\documentclass{solarphysics}

\usepackage[hyperref,optionalrh,solaromanenum]{spr-sola-addons}
\usepackage{solar_physics_macros}

\usepackage{graphicx}
\usepackage{amssymb,bm}
\usepackage{xcolor}

\newcommand{\ion}[2]{\textrm{#1}\,\textsc{\romannumeral #2}}

\newcommand{\rev}[1]{#1}

\begin{document}

\begin{article}

\begin{opening}

	\title{The \textit{Visible Spectro-Polarimeter} of the \textit{Daniel~K.~Inouye Solar Telescope}}
	\runningtitle{The \textit{Visible Spectro-Polarimeter} of the \textit{Daniel~K.~Inouye Solar Telescope}}
	\runningauthor{De Wijn, Casini, et al.}

	\author[addressref={aff1},corref,email={dwijn@ucar.edu}]{\inits{A.G.}\fnm{A.G.}~\lnm{de Wijn}\orcid{0000-0002-5084-4661}}
	\author[addressref={aff1},email={casini@ucar.edu}]{\inits{R.}\fnm{R.}~\lnm{Casini}\orcid{0000-0001-6990-513X}}
	\author[addressref={aff1}]{\inits{A.}\fnm{A.}~\lnm{Carlile}}
	\author[addressref={aff1}]{\inits{A.R.}\fnm{A.R.}~\lnm{Lecinski}}
	\author[addressref={aff1}]{\inits{S.}\fnm{S.}~\lnm{Sewell}\orcid{0000-0002-7252-4976}}
	\author[addressref={aff1}]{\inits{P.}\fnm{P.}~\lnm{Zmarzly}}
	\author[addressref={aff2}]{\inits{A.D.}\fnm{A.D.}~\lnm{Eigenbrot}\orcid{0000-0003-0810-4368}}
	\author[addressref={aff2}]{\inits{C.}\fnm{C.}~\lnm{Beck}\orcid{0000-0001-7706-4158}}
	\author[addressref={aff2}]{\inits{F.}\fnm{F.}~\lnm{W\"oger}}
	\author[addressref={aff1}]{\inits{M.}\fnm{M.}~\lnm{Kn\"olker}}

	\address[id=aff1]{High Altitude Observatory, National Center for Atmospheric Research, P.O.~Box 3000, Boulder, CO 80301-3000, USA.}
	\address[id=aff2]{National Solar Observatory, 3665 Discovery Dr., Boulder, CO 80301, USA}

	\begin{abstract}
		The \textit{Daniel K. Inouye Solar Telescope} (DKIST) \textit{Visible Spectro-Polarimeter} (ViSP) is a traditional slit-scanning spectrograph, with the ability to observe solar regions up to a \rev{$120\times78~\mathrm{arcsec}^2$ area}.
		The design implements dual-beam polarimetry, a polychromatic polarization modulator, a high-dispersion echelle grating, and three spectral channels that can be automatically positioned.
		A defining feature of the instrument is its capability to tune anywhere within the 380\,--\,900~nm range of the solar spectrum, allowing for a virtually infinite number of combinations of three wavelengths to be observed simultaneously.
		This enables the ViSP user to pursue well-established spectro-polarimetric studies of the magnetic structure and plasma dynamics of the solar atmosphere, as well as completely novel investigations of the solar spectrum.
		Within the suite of first-generation instruments at the DKIST, ViSP is the only wavelength-versatile spectro-polarimeter available to the scientific community.
		It was specifically designed to be a discovery instrument, for the exploration of new spectroscopic and polarimetric diagnostics, and to test improved models of polarized line formation, through high \rev{spatial-, spectral-, and temporal-resolution} observations of the Sun's polarized spectrum.
		In this instrument article, we describe the science requirements and design drivers of ViSP, and we present preliminary science data collected during the commissioning of the instrument.
	\end{abstract}


\end{opening}

\section{Introduction}

The \textit{Daniel K. Inouye Solar Telescope} \citep[DKIST:][]{2020SoPh..295..172R} is the first of a new generation of large ($> 2$~m) solar telescopes.
Its 4-m aperture will enable the solar-science community to undertake observations of structures in the solar atmosphere at an unprecedented spatial resolution ($<20~\mathrm{km}$ at the shortest wavelengths), and at the same time it provides the largest photon collecting area available for solar studies, creating exciting opportunities for the exploration of new spectral diagnostics and weakly polarized signals on the Sun.

The DKIST is designed with a facility suite of instruments with the capability and flexibility to address a large number of scientific questions, and, in the long term, it allows for adaptability to new scientific challenges.
Four out of five of its ``first-light'' instruments are spectro-polarimeters, aimed at the study of the different polarization signatures of the solar magnetic field over a broad range of spatial and temporal scales, and a large interval of optical to mid-\rev{IR} wavelengths (380 to 5000~nm).
The DKIST positions itself at the forefront of \rev{solar} science with this suite of instrumentation: spectro-polarimetry is a powerful tool for quantitative diagnostics of magnetic field in the solar \rev{plasma}, which is critically important for our understanding of physical processes in the solar atmosphere.
The \textit{Visible Spectro-Polarimeter} (ViSP) described in this article is one of these first-light instruments.

ViSP is a spectro-polarimeter based on a slit-scanning echelle spectrograph, designed to perform polarization measurements of solar radiation in the visible and near-\rev{IR} spectrum from 380 to 900~nm.
ViSP can observe up to three spectral regions at once with its three camera arms, enabling powerful multi-line diagnostics that probe plasma parameters in many different physical layers of the solar atmosphere simultaneously.
The wavelength band accessible to ViSP is determined by the DKIST \textit{Facility Instrument Distribution Optics} (FIDO) that feeds light to ViSP and other instruments.
FIDO consists of a series of mirrors, windows, and dichroic beam splitters that can be reconfigured to direct a single continuous region of the solar spectrum to an instrument \rev{\citep{2021JATIS...7d8005H}}.

ViSP is ``wavelength versatile'', meaning that it can be set up to observe \emph{any} wavelength within its spectral range of operation, while simultaneously achieving high spectral, spatial, and temporal resolution.
It is the only instrument at the DKIST with this capability at this time.
Hence, ViSP is not only ideally suited for multi-line diagnostics, but it is also a highly capable \emph{research} spectro-polarimeter that can be used to study any region of the solar spectrum within its wavelength range.

Most modern solar telescopes are \rev{equipped} with spectro-polarimeters \citep{2019OptEn..58h2417I}.
Imaging spectro-polarimeters using \rev{Fabry--P\'erot} interferometers have become commonly available at many telescopes, such as the \rev{\textit{Triple Etalon SOlar Spectrometer}} \citep{1998A&A...340..569K,2002SoPh..211...17T} at the \rev{\textit{German Vacuum Tower Telescope} \citep[VTT:][]{1998NewAR..42..493V}}, and the \rev{\textit{CRisp Imaging SpectroPolarimeter} \citep{2008ApJ...689L..69S}} at the \rev{\textit{Swedish 1-m Solar Telescope} \citep[SST:][]{2003SPIE.4853..341S}}.
These instruments have the benefit that they simultaneously capture a 2D region of the Sun, but they must scan in wavelength, potentially resulting in spectra that are not temporally coherent.
They are also restricted in their ability to tune in wavelength due to their limited free spectral range, which requires the use of a dedicated pre-filter for each wavelength region that is to be observed.

Only a few instruments with capabilities similar to ViSP currently exist.
Most spectrograph-based polarimeters such as the \rev{\textit{Polarimetric Littrow Spectrograph}} at the German VTT \citep{2005A&A...437.1159B}, the \rev{\textit{Facility Infrared Spectropolarimeter}} at the \rev{\textit{Dunn Solar Telescope} \citep[DST:][]{2010MmSAI..81..763J}}, the \rev{\textit{GREGOR Infrared Spectrograph}} \citep{2012AN....333..872C}, and the \rev{\textit{Hinode} \textit{Spectro-Polarimeter}} \citep{2013SoPh..283..579L} do not provide the same amount of flexibility and wavelength channels as ViSP.
The most similar is the \rev{\textit{Spectro-Polarimeter for Infrared and Optical Regions}} (SPINOR) at the DST \citep{2006SoPh..235...55S}, which replaced the \rev{\textit{Advanced Stokes Polarimeter}} \citep{1992SPIE.1746...22E}.
The echelle spectrograph at the German VTT and the \rev{\textit{TRI-Port Polarimetric Echelle-Littrow spectrograph}} at the SST \citep{2011A&A...535A..14K} are also similar but are not typically used as polarimeters.
ViSP improves on these instruments not only in terms of spatial resolution, but also in its automated configuration of grating and camera arm angles, significantly reducing setup time while improving repeatability.

\section{Science Objectives}

ViSP provides measurements of the full state of polarization simultaneously in up to three wavelength regions within the visible and near-\rev{IR} solar spectrum, with high polarimetric precision and high spectral resolving power.
Such measurements provide quantitative diagnostics of the magnetic field vector as a function of height in the solar atmosphere, along with the associated variation of the thermodynamic properties, that are crucial to addressing many of the DKIST science use cases \citep{rimmele2005atst}, such as improving our understanding of solar magnetism and of the trigger mechanisms of solar energetic events (i.e., flares and coronal mass ejections) that are the main causes of \rev{space weather}.

The spectral versatility and continuous wavelength coverage of ViSP are primary requirements, and two of the main drivers of the instrument design.
They are specifically aimed at making the instrument a research tool of polarization diagnostics over the entire visible solar spectrum.

A few key topics of investigations that can be performed with ViSP are:
\begin{itemize}
	\item the evolution of small-scale magnetism in the photosphere, including the quiet Sun, \rev{active-region} plage, and coronal holes;
	\item the emergence, evolution, and decay of active regions, and the fine structure of sunspots;
	\item the precursors and triggers of solar flares and coronal mass ejections, including filaments and prominences;
	\item the mass and energy cycle in the lower solar atmosphere and into the corona;
	\item the connectivity of the non-eruptive solar atmosphere through magnetic field; and
	\item oscillations in the photosphere and chromosphere.
\end{itemize}
A comprehensive review of the science objectives of ViSP is out of scope for this article.
We refer the interested reader to \cite{2021SoPh..296...70R} for details on the \textit{DKIST Critical Science Plan} that will be executed in the first years of operation.

\rev{Polarized light emerges as a manifestation of symmetry breaking processes that affect the interaction of radiation with matter (e.g., magnetic and electric fields, velocity field and thermal gradients, plasma wave processes, radiation scattering in the higher atmospheric layers).
Therefore, the ability to measure the polarized} solar spectrum is key to unveil the fundamental physical mechanisms driving the solar atmosphere and its short- and long-term variability, because of the dominant role that \rev{such processes of} plasma and radiation anisotropy play in the manifestation of the observed solar phenomena.
The ability to simultaneously tune a spectro-polarimeter on different spectral diagnostics serves a double purpose: first, lines with different formation heights can be used to produce a ``tomography'' of the solar atmosphere; second, the redundancy offered by observing multiple, independent spectral lines with different properties that are known to sample approximately the same height, can be of great help in reducing data analysis and inversion model errors and degeneracy \citep[see, e.g.,][]{1973SoPh...32...41S,2008ApJ...674..596S,2012SoPh..280..355B,2012SoPh..276...43D,2021A&A...653A.165K}.
The polarization signatures that are typically observed in the magnetized solar atmosphere result from a variety of effects, the most commonly known of which are the \emph{Zeeman} and \emph{Hanle effects}.

The Zeeman effect \citep{1897ApJ.....5..332Z,1897Natur..55..347Z} produces a wavelength splitting of magnetic-sensitive spectral lines (with non-zero Land\'e factor), which can then be detected and quantified through polarimetric observations with sufficiently high spectral resolution.
Most of the magnetic diagnostics of the photosphere is based on the interpretation of Zeeman-effect signatures in spectral lines such as the \ion{Fe}{1} 630.2~nm doublet, or the \ion{Fe}{1} 524.7 and 525.0~nm line pair.

Radiation scattering becomes a dominant process in the formation of polarized spectra when the plasma density and collision rates drop.
Plasma parameters depart strongly from local thermodynamic equilibrium (LTE) under these conditions, greatly increasing the complexity of the modeling and the difficulty of the interpretation of observations \rev{\citep[see, e.g., the review by][]{2020LRSP...17....3L}}.
This is the typical situation for \rev{spectral} diagnostics of the chromosphere, especially at the line cores.
Another characteristic of non-LTE line formation is the importance of the anisotropy of the solar radiation due to the presence of density and thermal gradients.
The most notable example is the phenomenon of limb darkening or brightening caused by the variation of radiation temperature with the solar radius.

Anisotropic excitation of atomic systems leads to the appearance of atomic \rev{sub}-level population imbalances, which produces a naturally polarized \rev{scattering} of the incident radiation.
\rev{As the different sub-levels de-excite, the atom emits radiation carrying imprints of the initial anisotropy \citep[see, e.g.,][]{2004ASSL..307.....L}}.
This mechanism is also present in the photosphere, but there it is greatly reduced because the larger collision rates at photospheric densities effectively \rev{equalize (i.e., ``thermalize'')} the atomic \rev{sub-level} populations.
However, even in the photosphere there are a few exceptions, such as the scattering dominated photospheric line of \ion{Sr}{1} at 460.7~nm all over the disk \citep[e.g.,][]{2018A&A...614A..89B}, or when observing the photosphere at the extreme solar limb \citep{2010ApJ...713..450L}.
In the chromosphere, instead, practically all spectral lines show the polarization signature of anisotropic excitation, which can be observed even in the absence of magnetic fields, \rev{especially close to the solar limb}.
The \rev{degree of linear polarization of the solar spectrum near the limb \citep[the ``second solar spectrum,'' e.g.,][]{1997A&A...321..927S}} demonstrates very well the great complexity of the polarization pattern produced under this mechanism.

For these scattering-dominated lines, when a magnetic field is also present, the linear polarization \rev{may be} modified, through a physical mechanism known as the Hanle effect \citep{1924ZPhy...30...93H}.
By means of high-sensitivity observations that can detect the polarization changes with respect to an ideal model of the field-free plasma, it then becomes possible to diagnose magnetic fields that are much weaker than the ones accessible to the Zeeman-effect diagnostics \citep[see, e.g.,][]{1982SoPh...80..209S}.
In addition, the Hanle effect is much less sensitive than the Zeeman effect to polarization cancellation in the presence of unresolved magnetic structures \citep[such as in turbulent local dynamo; e.g.,][]{2009ApJ...693.1728P}.
Like in the application of the Zeeman effect, \rev{the} use of multiple \rev{spectral} diagnostics is also very important for the Hanle effect, especially because of its reliance on atmospheric modeling for setting the polarization reference level corresponding to the ideal field-free atmosphere \citep{1998A&A...329..319S}.

Atomic-level interference and crossings, induced by magnetic fields in complex atoms, produce even more exotic polarization effects, such as the alignment-to-orientation (A-O) mechanism responsible for the appearance of intensity-like signals in Stokes $V$ \rev{\citep{2004ASSL..307.....L}}.
These greatly enrich the diagnostic potential of scattering lines.
In particular, the vector field structure within prominences as determined from the Hanle effect and the A-O mechanism is an objective of the critical science plan \citep{2021SoPh..296...70R}.
Such an analysis requires high sensitivity to both linear and circular polarization because of the subtlety of these effects, and it also benefits from the simultaneous observations of multiple diagnostics.

\section{Requirements}

The science objectives described above drive ViSP instrument requirements.
The main two drivers of the instrument design are multi-line diagnostics and spectral versatility, i.e., the requirement to observe up to three spectral regions simultaneously and the ability to observe any line in the visible and near IR (\rev{380} to 900~nm).

Multi-line diagnostics are key for providing a ``snapshot'' of the thermodynamic and magnetic structure of the solar atmosphere, owing to the fact that different spectral lines generally have different ranges of temperatures and formation heights in the atmosphere.
Accordingly, ViSP has a requirement to observe three wavelength regions simultaneously.
The wavelength range of ViSP includes magnetic and plasma diagnostics that span from the lower photosphere to the low corona. Thus, plasma temperatures accessible to ViSP reach down to the chromospheric temperature minimum, while the few coronal emission lines in ViSP range are sensitive to temperatures ranging approximately from 1.1 to 2~MK.
However, no spectral diagnostics probing the solar transition region between the chromosphere and the low corona are accessible to ViSP or DKIST, as these all belong to the UV spectrum that lies below the \rev{terrestrial} atmosphere's wavelength cutoff.

ViSP is intended to also serve as an exploratory instrument for the whole polarized solar spectrum in the visible and near-IR.
For this reason, it must be capable of observing combinations of spectral lines regardless of whether these belong to widely different parts of the solar spectrum or to the same spectral region, e.g., close multiplets such as the \ion{Ca}{2} H and~K lines (\rev{$393$ and $397~\mathrm{nm}$}), the \ion{Mg}{1} b1-b3 lines (\rev{around} $517~\mathrm{nm}$), the \ion{Na}{1} D doublet (\rev{around $589~\mathrm{nm}$}), and the \ion{Ca}{2} IR triplet (\rev{$850$, $854$, and $866~\mathrm{nm}$}).

\renewcommand{\arraystretch}{1.25}
\begin{table}[tbp]
	\begin{tabular}{l l l}
		\hline
		\textbf{Parameters}         & \textbf{Requirement}              & \textbf{Notes} \\
		\hline
		Wavelength range            & 380\,--\,900~nm                   & \\
		Simultaneous wavelengths    & \rev{three} lines                 & \\
		Spectral resolving power    & 180\,000                          & \\
		Spatial FOV                 & $120\times120~\mathrm{arcsec}^2$  & \parbox[t]{1.25in}{limited by cameras\\ to $<120\times78~\mathrm{arcsec}^2$} \\
		Spatial resolution          & \parbox[t]{1.25in}{$2\times$ DKIST resolution at all wavelengths}
									& \parbox[t]{1.3in}{\rev{met for $\lambda\ge450~\mathrm{nm}$ due\\ to
									the width of the\\ narrowest slit}} \\
		Slit scan repeatability     & $\pm\frac{1}{2}$ slit width       & measured $<1\%$ slit width \\
		Slit scan accuracy          & $0.1~\mathrm{arcsec}$             & measured $0.03~\mathrm{arcsec}$ \\
		Polarimetric sensitivity    & $10^{-4}\,I_\mathrm{cont}$        & \\
		Polarimetric accuracy       & $5\times10^{-4}\,I_\mathrm{cont}$ & \\
		Temporal resolution         & \parbox[t]{1.25in}{\rev{10~s to reach} $10^{-3}\,I_\mathrm{cont}$\\ for $\lambda>500~\mathrm{nm}$}
									& \parbox[t]{1.25in}{measured $10^{-3}\,I_\mathrm{cont}$ in 2~s at 630~nm} \\
		Spectral bandpass           & 1.1~nm at 630~nm                  & \\
		Setup time                  & 1 channel in 10~min               & \\
		Slit move time              & \parbox[t]{1.25in}{200~ms between adjacent\\ slit positions}
									& \\
		Slit slew velocity          & 2~arcmin in 30~s                & \\
		\hline
	\end{tabular}
	\caption{Summary of science and operational requirements of ViSP.}
	\label{tab:reqs}
\end{table}
\renewcommand{\arraystretch}{1}

The principal science requirements of ViSP concern performance in the spectral, spatial, and temporal domains.
\rev{These requirements were derived from the DKIST science use cases \citep{rimmele2005atst}.
The spectral range is driven by the desire to observe the \ion{Ca}{2} lines at 393.4 and 866.2~nm.}
ViSP is designed to attain high spectral resolution ($R \gtrsim 180\,000$).
The optical design is diffraction limited over the DKIST spectral range where the adaptive optics system can effectively correct for seeing aberrations (approximately $\lambda \gtrsim 500$~nm), exceeding its requirement to meet the \rev{spatial-resolution} requirement of \rev{two} times the DKIST diffraction limit.
The instrument throughput is optimized to achieve a polarimetric sensitivity of $10^{-3}\,I_\mathrm{cont}$ within \rev{ten} seconds of signal integration in the region of the solar spectrum with $\lambda \gtrsim 500$~nm.
In addition, ViSP is outfitted with three automatically reconfigurable camera arms in order to observe simultaneously in three different spectral passbands.
A list summarizing the science and operational requirements for ViSP is given in Table~\ref{tab:reqs}.

\section{Instrument Design}\label{sec:design}

\begin{figure}[tbp]
	\centering
	\includegraphics[width=1.\textwidth]{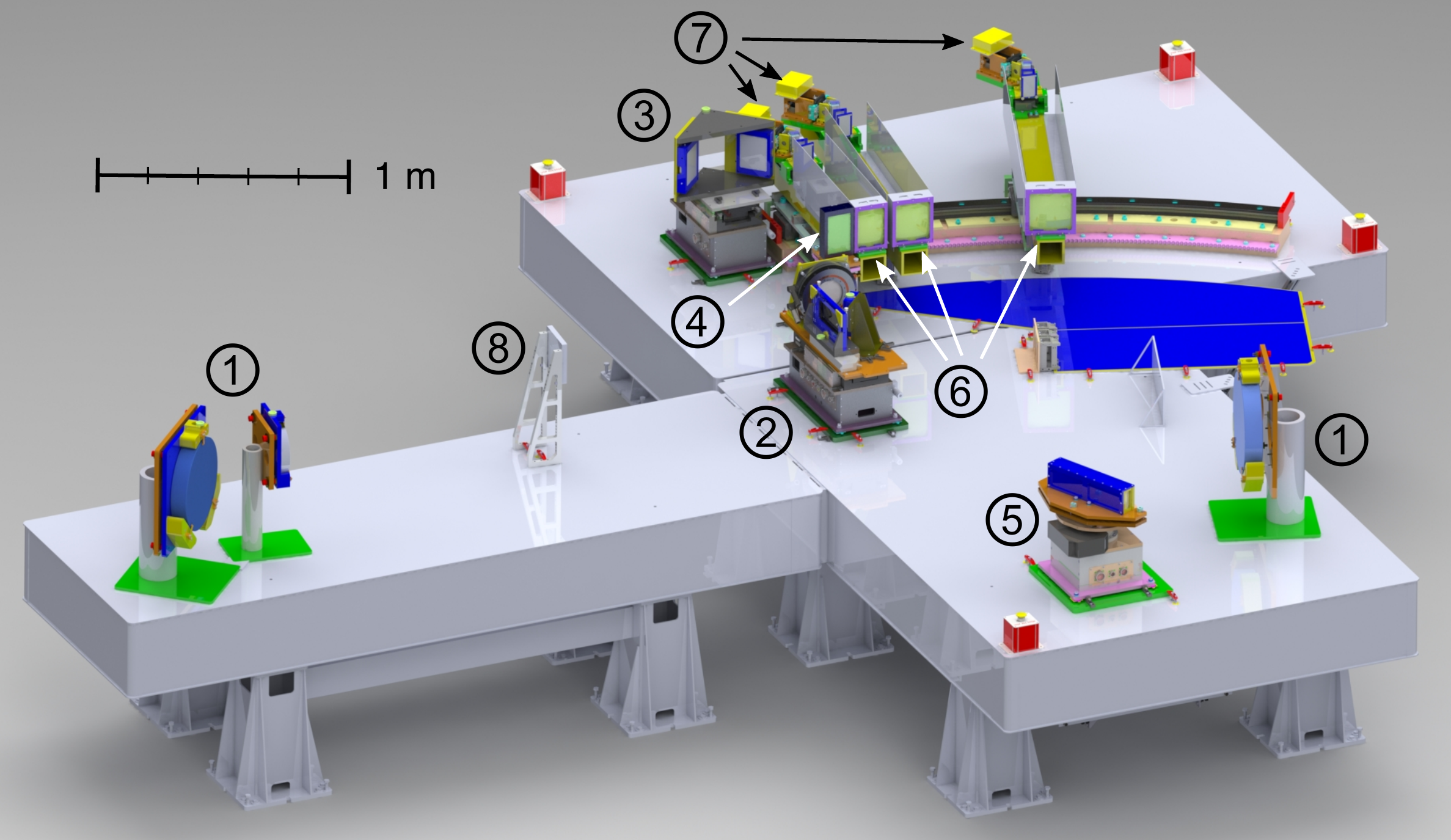}\vspace{6pt}
	\includegraphics[width=1.\textwidth]{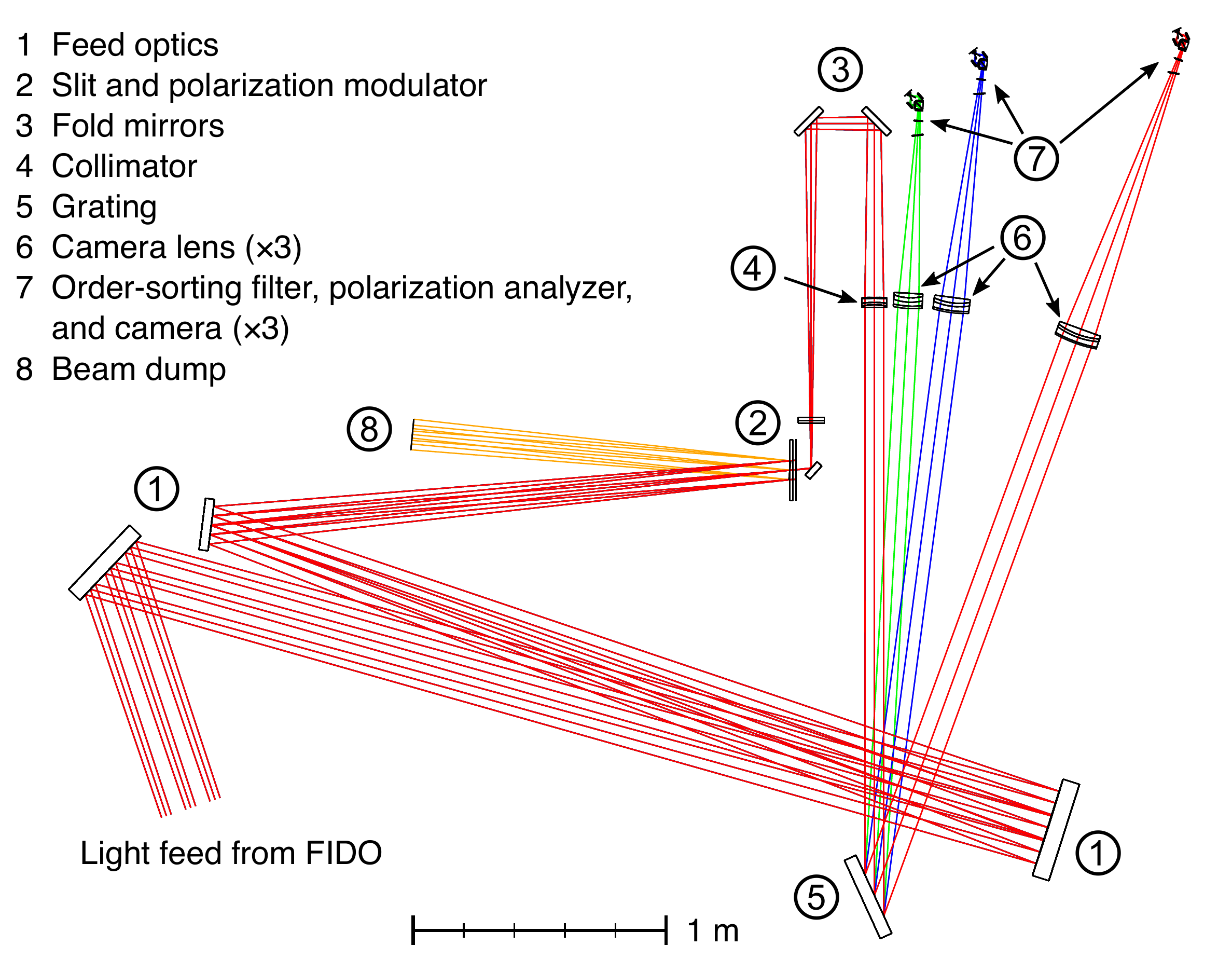}
	\caption{%
		\rev{\textit{Top:}} 3D-model rendering of ViSP instrument on its optical tables.
	\rev{\textit{Bottom:}} top-down view of the optical layout of the instrument in the Science Verification configuration with components labeled.}
	\label{fig:model}
\end{figure}

The main challenge in designing a wavelength-versatile spectro-polarimeter such as ViSP is that it is not possible to meet the large set of science-driven instrument requirements (spatial resolution, field of view, spectral resolution, bandwidth, instrument throughput) simultaneously for all possible combinations of spectral lines.
Therefore, the requirements must be ranked based on scientific priority balanced with technical feasibility and cost of fabrication and implementation.

The starting point of the ViSP design was to identify the constraints for the dimensions of the instrument.
This is a sensible approach, since the complexity of the design and the cost of fabrication are strongly dependent on the size of the instrument and particularly the optics.
ViSP consists of a number of opto-mechanical components and systems set up on three joined optical tables.
Renderings of the ViSP mechanical optical and mechanical models are shown in \rev{Figure}~\ref{fig:model}.

\subsection{\rev{Feed Optics and Slit Focal Plane}}\label{sec:feedopt}

The requirement for ViSP to reach a spatial resolution of \rev{two} times the diffraction limit of the DKIST at all visible wavelengths implies that the entrance slit must be able to critically sample the resolution element down to 380~nm (i.e., 0.024~arcsec for a 4-m aperture).
At the same time, the requirement for a spatial field of view (FOV) of $2{\times}2~\mathrm{arcmin}^2$ forces the entrance slit to have a very large aspect ratio (\rev{$\approx5000{:}1$}).
Because the slit length is constrained to keep the focal plane to a reasonable size, the slit must be narrow, in the range of $10$ to $20~\mathrm{\mu m}$ wide.
Such a slit can be fabricated by photo-lithography with good tolerances ($\pm1~\mathrm{\mu m}$) over the full length of the slit of several cm.
For the DKIST, this implies an effective focal length \rev{$(f/\#)_\mathrm{tel}\approx30$} at the spectrograph entrance.

The feed-optic (FO) system of ViSP is a modified ``Schiefspiegler'' telescope delivering \rev{$(f/\#)_\mathrm{tel}\approx32$} (see items labeled~1 in \rev{Figure}~\ref{fig:model}).
It consists of three spherical mirrors, although the first mirror is nearly flat and mainly acts as a beam-steering optic.
The choice of this design was driven by its simplicity and \rev{cost-effectiveness}, and the good aberration control attainable for relatively large $f$-numbers.
The ViSP FO system is able to produce an image of the $2{\times}2~\mathrm{arcmin}^2$ FOV that is both flat (within the depth of focus of the collimator) and highly telecentric.
This allows the instrument to preserve the image focus at the slit and a constant illumination of the grating across the entire FOV.
With this FO system, the \rev{critical-sampling} width at the focal plane is \rev{about} $15~\mathrm{\mu m}$ at 380~nm, with a slit length of \rev{about} $75~\mathrm{mm}$ (i.e., a plate scale of \rev{approximately $1.6~\mathrm{arcsec}\,\mathrm{mm}^{-1}$}).
The off-axis design of the FO telescope produces a focal plane that is intentionally tilted by $5.4^\circ$.
This allows the unwanted light reaching the reflective slit optic to be rejected towards a beam dump, rather than back into the optical path of the incoming beam, \rev{and also allows for the possible addition of a context imager.}

ViSP is equipped with a \rev{set} of \rev{five} slit apertures, photo-etched on a single \rev{optical-glass} support with an aluminium reflective coating (\rev{Item}~2 in \rev{Figure}~\ref{fig:model}).
The first three slit apertures have widths that match the diffraction-limit resolution of the DKIST at 450, 650, and 850~nm (respectively sampling at 0.028, 0.041, and 0.053~arcsec), hence somewhat compromising the \rev{spatial-resolution} requirement towards the blue end of the spectrum.
A fourth and fifth slit, respectively sampling at approximately 0.1 and 0.2~arcsec, are included in order to address use cases where it is necessary for the science to trade spectral and spatial resolution for high throughput (e.g., for prominence and coronal observations).
Fiducial hairlines with a separation of 45.2~arcsec are directly etched into the substrate for alignment purposes.

Field scanning is achieved by moving the slit on an Aerotech ANT180-260-L translation stage.
The slit optic is mounted on an Aerotech MPS75SLE stage.
The slit is selected by moving the slit optic in front of a mask that allows light from only one slit to pass into the spectrograph while also helping to reduce stray light from pinholes in the reflective coating of the slit optic.

\subsection{Spectrograph}\label{sec:spectro}

The spectrograph consists of a collimator lens (\rev{Item}~4 in \rev{Figure}~\ref{fig:model}), which forms an image of the pupil of the telescope on the diffraction grating, the grating proper (\rev{Item}~5 in \rev{Figure}~\ref{fig:model}), and a camera telescope for each of the spectral channels.
These systems are all interconnected, and the characteristics of one constrain the design of the other.
Here, we discuss first the selection of the diffraction grating, and then \rev{we} turn our attention to the collimator, and finally the camera arms.

\subsubsection{Diffraction \rev{Grating}}\label{sec:grating}

The use of a commercial off-the-shelf (COTS) grating was forced by cost and schedule constraints.
Thus, we had to ensure that the selected grating allowed us to meet the spectral resolving power requirement while also enabling the required spectral versatility.

\begin{figure}[tbp]
	\centering
	\includegraphics[width=4in]{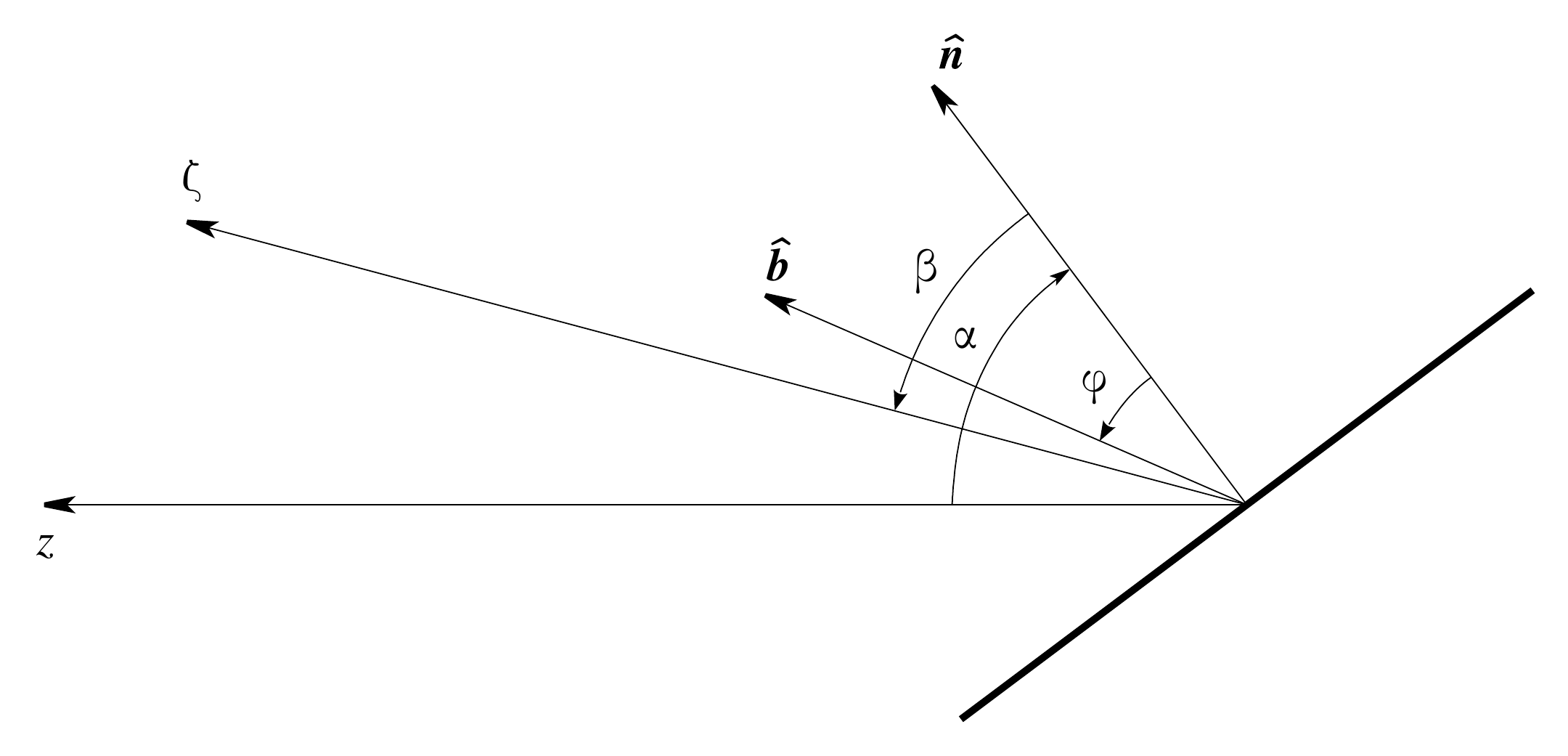}
	\caption{%
		Geometric convention adopted in this work for the definition of the incidence and diffraction angles (respectively, $\alpha$ and $\beta$) relative to the grating normal \rev{$\bm{\hat n}$}.
		The direction of the incident beam is $-z$, whereas that for the diffracted beam is $+\zeta$.
		The deviation angle between the diffracted and incident beams is defined by $\delta=\alpha+\beta$, and \rev{it} is a negative quantity for the normal operation of ViSP (assuming the usual convention of counterclockwise angles being positive).
	The blaze angle $\varphi$ measures the inclination of the normal to the blazed facets of the grating, \rev{$\bm{\hat b}$}, from the grating normal.}
	\label{fig:layout}
\end{figure}

The spectral resolution of the instrument depends, among other things, on the \emph{finesse} profile of the grating \cite[e.g.,][]{CN14}
\begin{equation}
	\mathcal{F}(\alpha,\beta)=\mathrm{sinc}^2\biggl[\pi\,
	\frac{L}{\lambda}\,(\sin\beta-\sin\alpha)\biggr],
\end{equation}
where $\alpha$ and $\beta$ are, respectively, the incidence angle of the incoming beam on the grating and the emergence angle of the diffracted beam, measured from the grating normal (see \rev{Figure}~\ref{fig:layout} for the diffraction geometry and sign convention for those angles), and $L$ is the illuminated width of the grating.
In order to achieve a target resolving power $R$ at a given wavelength and incidence angle $\alpha$, the corresponding dispersion interval,
\begin{equation}\label{eq:dispers}
	\delta\beta_R=\frac{1}{R}\,\frac{\sin\beta-\sin\alpha}{\cos\beta},
\end{equation}
must be critically sampled by the full width at half maximum (FWHM) of the finesse profile as a function of $\beta$,
\begin{equation}\label{eq:finesse}
	\delta\beta_\mathrm{g}=\frac{\lambda}{L\cos\beta},
\end{equation}
i.e., \rev{$\delta\beta_R\approx2\,\delta\beta_\mathrm{g}$}.
In reality, one must also consider the finite width of the slit, which contributes to the line-spread function of the spectrograph along with $\delta\beta_\mathrm{g}$ \citep[see][]{CdW14}.
Considering for simplicity the case where the spectrograph is set to the Littrow configuration and with the grating tilted at the blaze angle ($\beta=-\alpha=\varphi$), such a sampling condition gives
\begin{equation}\label{eq:Rset}
	\rev{R\approx\frac{L}{\lambda}\sin\varphi=\frac{w_\mathrm{C}}{\lambda}\tan\varphi},
\end{equation}
where $w_\mathrm{C}$ is the projected width of the grating on the plane normal to the optical axis of the collimator.
This must at least match the width of the collimator optic to avoid vignetting by the grating.
We then see that the larger the blaze angle $\varphi$ the narrower the collimator and grating optics can be made, decreasing fabrication risk and cost.
On the other hand, too large values of the blaze angle practically limit the range of tilt angles at which the grating can be used, impacting the spectral versatility of the instrument.
A collimator width of \rev{$w_\mathrm{C}\approx10~\mathrm{cm}$} (see \rev{Section}~\ref{sec:collim}) implies the use of gratings with blaze angles \rev{$\varphi\approx60^\circ$} and \rev{$L\approx20~\mathrm{cm}$} in order to meet the \rev{spectral-resolution} requirement.

\rev{It is important to remark that meeting the spectral-resolution requirement is subject to the condition that the effective width of the grating be coherently illuminated.
This is usually achieved in ViSP, e.g., when using a wavelength-scaled slit (see \rev{Section}~\ref{sec:collim}), as this coherently samples the FOV along the spectral dimension \cite[a consequence of the Van \rev{Cittert--Zernike} theorem; e.g.,][]{MW95}.}

\begin{figure}[tbp]
	\centering
	\includegraphics[width=\textwidth]{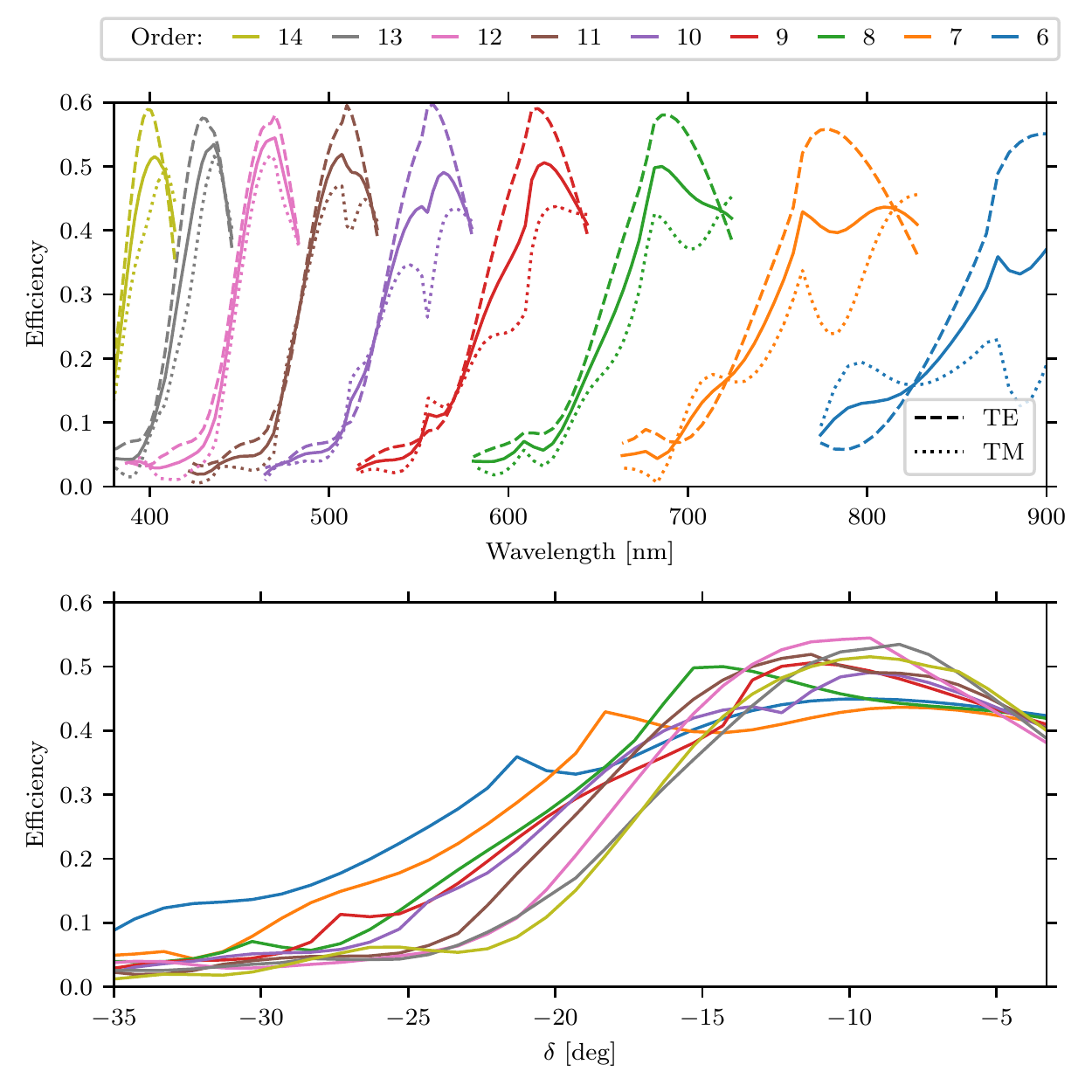}
	\caption{Model of the vector (polarized) efficiency of the ViSP grating for a tilt angle $\alpha=-68^\circ$, working between the diffraction orders~6 and 14, and for deviation angles $\delta=\alpha+\beta$ between $-3.3^\circ$ and $-35.3^\circ$.
		\textit{Top}: the two polarized efficiencies TE (groove-parallel, \rev{\textit{dashed curves}}) and TM (groove-perpendicular, \rev{\textit{dotted curves}}), along with the corresponding average (\rev{\textit{solid}} curves).
	\textit{Bottom}: the average efficiency curves plotted against the diffraction angle $\delta$.}
	\label{fig:grateff}
\end{figure}

Gratings have a characteristic \rev{efficiency curve for the diffracted energy, which determines the usable range of diffraction angles of a given spectrograph configuration (see \rev{Figure}~\ref{fig:grateff}, bottom panel)}.
A spectrograph with reconfigurable arms like ViSP must thus be capable \rev{of positioning} all the spectral channels used for an observation within this optimal angular range.
Ultimately this is a function of the groove spacing $d$ of the grating and the configuration of the spectrograph.
Using the scalar theory of diffraction, the usable range of diffraction angles for a given grating can be estimated by the FWHM of the \rev{grating-efficiency} curve \cite[e.g.,][]{CN14}\rev{:}
\begin{equation}\label{eq:efficiency}
	I(\alpha,\beta)=\mathrm{sinc}^2\biggl[\pi\,
		\frac{d}{\lambda}\,\cos\alpha\,\frac{\sin(\beta-\varphi)-\sin(\alpha+\varphi)}{\cos(\alpha+\varphi)}\,
	\biggr].
\end{equation}
We thus find that the \rev{range of usable $\beta$ angles} of a grating is approximately
\begin{equation}
	\Delta\beta_\mathrm{g}=\frac{\lambda}{d\cos\alpha}\,\frac{\cos(\alpha+\varphi)}{\cos(\beta-\varphi)}.
\end{equation}
In the case of ViSP, in order to access many scientifically interesting combinations of spectral lines with good efficiency, we must employ gratings that can deliver \rev{$\Delta\beta_\mathrm{g}\approx20^\circ$}.
Considering again the case of a Littrow configuration, with the grating tilted at the blaze angle ($\beta=-\alpha=\varphi$), this implies groove densities larger than \rev{about $300~l\,\mathrm{mm}^{-1}$} (e.g., for \rev{$\lambda\approx500~\mathrm{nm}$} and \rev{$\varphi\approx60^\circ$}).
Such gratings also have the advantage \rev{of working} in relatively low diffraction orders ($m\lesssim 15$) over the visible spectrum, allowing us to employ a correspondingly small number of order-sorting filters to fully cover the ViSP spectral range.

For ViSP, we adopted an \rev{aluminum}-coated echelle grating by Newport-RGL with a groove density of $316~l\,\mathrm{mm}^{-1}$, blazed at $63.4^\circ$, of $90\times340~\mathrm{mm}^2$.
An example of the polarized efficiency of the ViSP grating, based on a vector model of diffraction \citep{Li99}, is given in \rev{Figure}~\ref{fig:grateff}, for $\alpha=-68^\circ$ and a range of arm angles \rev{$\delta$} corresponding to the opto-mechanical capabilities of the instrument.

The grating is mounted on a Newport RV120HAHLT-F precision rotation stage for automated positioning.
While ViSP is outfitted with a single grating at this time, the mount uses an intermediate kinematic platform to facilitate manual exchange of gratings should more gratings become available for use in the future.

\subsubsection{Collimator}\label{sec:collim}

\rev{An oversized collimator optic is necessary} to avoid loss of instrument throughput as a result of diffraction, because of the matching condition of the \rev{entrance-slit} width to the diffraction limit of the telescope.
A rigorous diffraction-model analysis of \rev{telescope--spectrograph} systems \citep{CdW14} shows that, by making the collimator system faster than the imaging telescope by \rev{about} $20\%$ (i.e., \rev{$(f/\#)_\mathrm{tel}/(f/\#)_\mathrm{coll}\approx1.2$}, or \rev{$(f/\#)_\mathrm{coll}\approx27$}), the diffraction losses of ViSP remain below 10\%, when the entrance slit matches the diffraction limit of the DKIST at a given wavelength.

Recalling the argument leading to \rev{Equation}~(\ref{eq:Rset}), the focal length of the collimator is ultimately set by the width $w_\mathrm{C}$ necessary to meet the resolving power of $R\sim 180\,000$ at the detector for all wavelengths.
Using $\varphi=63.4^\circ$ and $\lambda=900~\mathrm{nm}$, we find $w_{\mathrm C}\gtrsim 9~\mathrm{cm}$ (also allowing for anamorphic magnification when $\alpha+\beta\ne0$; see~\rev{Equation}~(\ref{eq:dispers})).
We thus arrive at a focal length of \rev{about $2.4~\mathrm{m}$} for the collimator.

The ViSP collimator is an achromatic doublet with a focal length $f_\mathrm{coll}=2.37~\mathrm{m}$ and a clear aperture width of 90~mm.
The large $f/\#$ allows adequate control of aberrations using only spherical surfaces and a flat exit surface.
This facilitates the alignment of the instrument, as the flat surface can be used to reference an auto-collimating theodolite.
Controlling chromatic aberrations required the use of a combination of a low-dispersion fluorophosphate meniscus and a plano-convex borosilicate element.
The Ohara S-FPL53 and S-BSL7 glasses selected have substantially different coefficients of thermal expansion, which motivated the decision to fabricate the doublet with an air gap.
A slight improvement in performance could then be realized by allowing the internal radii to differ.
The collimator doublet was optimized not only for aberration control, but also for achromaticity of the pupil position, critical for control of spectrograph astigmatism at large \rev{arm} angles \rev{$\delta$}.

The \rev{width of the} collimator lens drives the physical width of the grating to be \rev{about} $33~\mathrm{cm}$, which allows configurations with tilt angles up to \rev{about} $74^\circ$ without suffering vignetting of the collimated beam by the grating.

Since diffraction effects can be neglected along the slit, the CA height of the collimator lens is only constrained by the size of the pupil (\rev{about} $74~\mathrm{mm}$) plus the height of the FOV at the slit plane (\rev{about} $75~\mathrm{mm}$).
Hence, it was set at $150~\mathrm{mm}$.
It is important to keep the horizontal extent of the collimator and the camera lenses to a minimum because they must be placed close together for spectral versatility.
The collimator doublet was therefore fabricated with a rectangular shape \rev{to match the shape of the beam}, and the mount was specifically designed to allow \rev{for Arm~1 to be positioned very close.}

\begin{figure}[tbp]
	\centering
	\includegraphics[width=\textwidth]{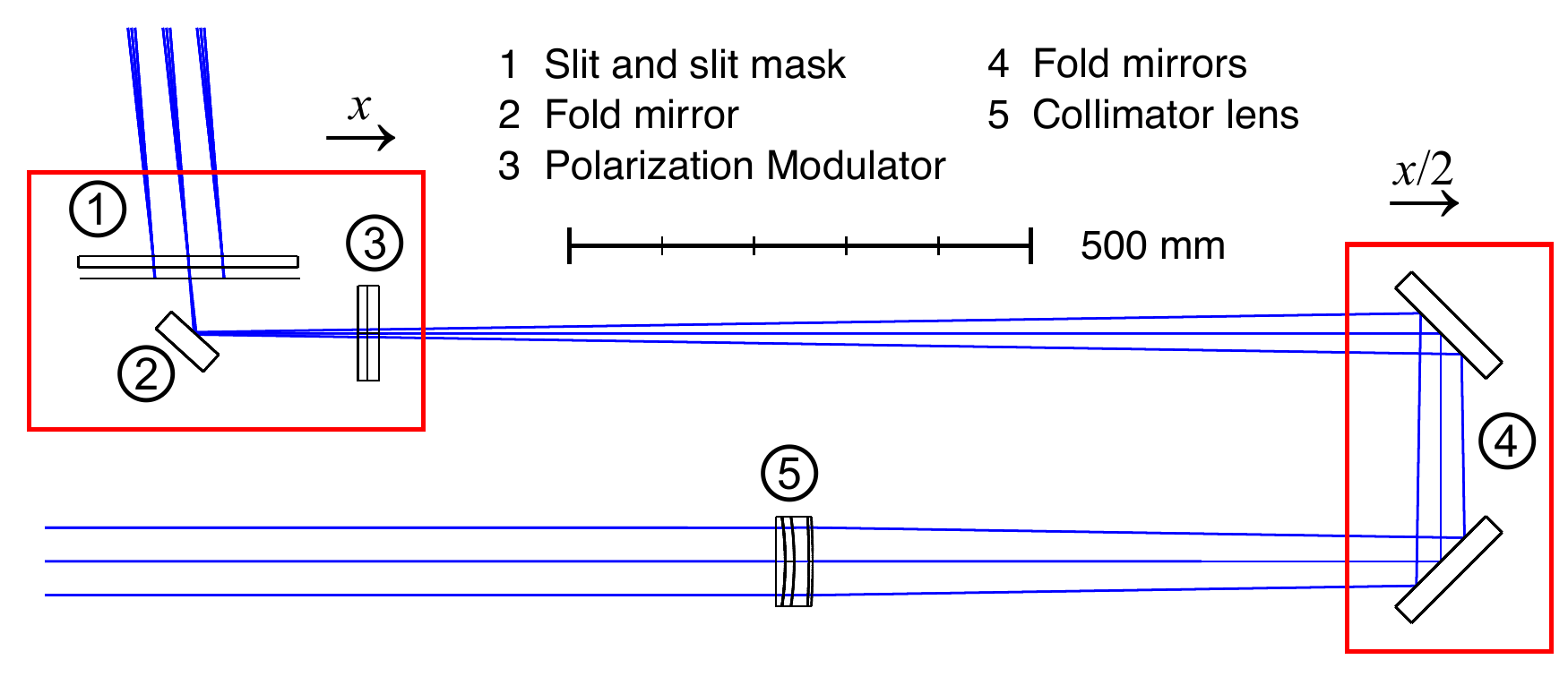}
	\caption{Optical ray tracing of the ViSP collimator.
		The image plane produced by the FO system is spatially filtered by the slit aperture on the slit optic (\textit{top left}).
		The input beam is deflected by a fold mirror into the rotating polarization modulator, and then folded backward by a pair of two mirrors at $90^\circ$ from each other.
		Finally, the beam passes through an achromat to be collimated onto the grating (not shown).
	In order to preserve the spectrograph focus during translation of the slit station by a distance $x$ in the slit-scanning direction, the path-folding station synchronously moves by $x/2$ in the same direction.}
	\label{fig:collimator}
\end{figure}

In order to keep the ViSP spectrograph compact, the collimator was designed with a folded optical path (see \rev{Figure}~\ref{fig:collimator}).
The fold mirror behind the slit directs the beam parallel to the slit image plane.
It and the polarimetric modulator are mounted with the slit optic on a single translation stage.
The focal distance between the collimator lens and the slit optic during the scanning of the FOV is preserved through a system of two fold mirrors at $90^\circ$ to each other (working as a \rev{retroreflector}), which must move synchronously with the slit motion at half the rate.
To this end, the two fold mirrors are mounted together on an Aerotech ANT180-160-L translation stage.
The instrument control software ensures that the motions are synchronized by commanding the motion-controller hardware to move the stages in unison.

\subsubsection{Spectral \rev{Channels}}\label{sec:cameras}

The collimator width determines also the width of the camera lenses in the three spectral channels.
Ideally, the width of a camera lens for a given spectral channel must take into account the anamorphic magnification, $r=\cos\alpha/\cos\beta$, pertaining to the particular configuration of that camera arm.
However, ViSP has reconfigurable camera arms and there are no preassigned values of the anamorphic magnification for any of the channels.

\begin{table}[tbp]
	\begin{tabular}{crrr}
		\hline
		\textbf{Spectral channel} & \textbf{\#1} & \textbf{\#2} & \textbf{\#3} \\
		\hline
		$\langle r\rangle^{-1}$ & 1.16 & 1.44 & 1.75 \\
		\hline
		FOV in arcsec $\lesssim$ & 78 & 62 & 52 \\
		\hline
	\end{tabular}
	\caption{Optimized scaling factors for the \rev{clear aperture} widths of the camera lenses relative to the \rev{clear aperture} width of the collimator \rev{and approximate fields of view}.}
	\label{tab:anamorph}
\end{table}

In order to optimize the camera-lens widths, we conducted a study of the distribution of spectrograph configurations for typical uses of the instrument, in order to identify the most ``probable'' positions (and corresponding anamorphic magnifications) of the three arms.
Such analysis led to the values for $\langle r\rangle^{-1}$ given in Table~\ref{tab:anamorph}.
These values were used to determine the CA widths of the camera lenses, and, like the collimator, these lenses were fabricated with a rectangular shape in order to maximize the spectral versatility of the instrument.

The focal lengths of the camera lenses are subject to both spatial and spectral constraints.
The requirement to capture the full FOV height of 120~arcsec on the detector sets the spectrograph magnification, and consequently the camera focal length.
The design assumed a detector with $6.5~\mathrm{\mu m}$ square pixels, yielding a spectrograph magnification \rev{$f_\mathrm{cam}/f_\mathrm{coll}\approx0.35$}, where $f_\mathrm{coll}$ and $f_\mathrm{cam}$ are the focal lengths of the collimator and the spectral channel, respectively.
The pixel size was based on the Andor Zyla 5.5 camera, in hopes that larger detectors of $4\mathrm{k}\times4\mathrm{k}$ pixels of the same size would become available.
The Andor Balor camera that is supported by DKIST and used by other instruments \citep[e.g., VBI\rev{:}][]{2021SoPh..296..145W} unfortunately has larger $12~\mathrm{\mu m}$ pixels.
It was determined that a \rev{redesign} with the larger pixel size would result in an unacceptable increase in cost and instrument size (due to the longer focal length of the camera lenses), and therefore the original design was implemented instead.
The Andor Zyla 5.5 cameras in ViSP are oriented to have 2560 pixels in the spatial dimension, and 2160 pixels in the spectral dimension.
Hence, the height of the captured FOV is at most 78~arcsec in \rev{Arm}~1.

It must be noted that the spectrograph magnification of \rev{Arm}~1 approximately maps the 450-nm slit width ($17.6~\mathrm{\mu m}$) to one pixel width.
This condition of ``pixel matching'' ensures that the FOV is critically sampled both along and transversely to the slit, in order to satisfy the \rev{spatial-resolution} requirement of twice the diffraction limit of the DKIST at least down to 450~nm in that channel.
Moreover, pixel matching is also key for ViSP to meet its \rev{spectral-resolution} target, i.e., to satisfy the sampling condition \rev{$\delta\beta_R\approx2\,\delta\beta_\mathrm{g}$} \rev{(see Section~\ref{sec:grating})}, and thus it must generally be implemented in all spectral channels.

In the spectral dimension, we must consider the effects of anamorphic magnification on the spectrograph \rev{point-spread} function \citep{CdW14}, which gets shortened in that dimension by the same factor\rev{:} $r=\cos\alpha/\cos\beta$. This required increasing the widths of the camera lenses in order to mitigate vignetting for configurations with $\beta<-\alpha$.
Therefore, in order to maintain as far as possible the ideal condition of pixel matching across the various configurations of the spectrograph, the focal lengths of the three spectral channels must also be scaled in the same proportions as given in Table~\ref{tab:anamorph}.
Hence, all camera lenses in ViSP were designed with the same \rev{$(f/\#)_\mathrm{cam}\approx8$} in the spectral dimension.
This allows us to remain close to pixel matching across the full range of diffraction angles.
On the other hand, it also implies that the height of the FOV that can be imaged onto the detector is proportionally smaller for \rev{Arms}~2 and~3 ($\lesssim 62~\mathrm{arcsec}$ and $\lesssim 52~\mathrm{arcsec}$, respectively), while their spatial resolution along the slit is correspondingly larger.

A side-effect of having different magnifications for the three spectral channels is the ability to trade \rev{FOV and bandwidth for resolution} by using different channels and diffraction angles to observe the same \rev{target}.

The collimator and camera lenses are placed at optically conjugate points with respect to the position of the system's pupil on the grating.
In this configuration, the collimated beam propagated via the grating reflection has the same height at the exit face of the collimator lens and the entrance face of the camera lenses.
Hence, the CA height of the camera lenses is set identical to that of the collimator.
The three camera lenses are cemented achromatic doublets.
Like the collimator, all surfaces are spherical and the exit surface is flat.
The three doublets all use the same combination of a high-dispersion lanthanide glass meniscus (Ohara S-LAL12), followed by a plano-convex borosilicate element (Ohara S-BSL7).
Since each arm observes only a small spectral window, it is possible to optimize the lenses for best monochromatic image quality while allowing for considerable chromatic shift in focal length over the ViSP spectral range.
The effect of this is more pronounced for lenses with longer focal lengths.
It amounts to \rev{about} $53~\mathrm{mm}$ of focal range in \rev{Channel}~3.
The polarization analyzer, the beam combiner optics, and the camera (see \rev{Sections}~\ref{sec:analyzer} and~\ref{sec:cameras}) are translated together on an Aerotech MPS75SLE stage to focus the image on the detector.
The arms are designed similarly, essentially differing only in length.
Similarly to the design of the collimator lens cell, the \rev{camera-lens} mounts are also designed for minimum width for highest spectral versatility.

The arms are mounted on independent carts that can be positioned with high precision along a section of THK HCR 65A curved rail with a radius of 3~m.
The carts are driven by Aerotech BMS60 motors through a Nexen HGP17 \rev{roller-pinion} drive system that is \rev{preloaded} into the curved rack to provide backlash-free operation.
The rail consists of two sections of $6$ and $30^\circ$ that were carefully aligned to be highly circular and placed so that its axis is in the plane of the grating surface.
The $6^\circ$ segment is placed on the side of the collimator, because modeling showed this results in fewer spectrograph configurations in which an arm spans the two rail segments.
The camera arms can move over a range of arm angles \rev{$\delta$} from $-3.3^\circ$ to $-35.3^\circ$, with a minimum separation of $3.55^\circ$ and $4.35^\circ$ between \rev{Arms}~1 and 2, and \rev{Arms}~2 and 3, respectively.

\subsubsection{Order-Sorting Filters}\label{sec:filters}

Because ViSP relies on the use of a diffraction grating working in multiple orders (see \rev{Section}~\ref{sec:grating}), each spectral channel must be equipped with an order-sorting filter to prevent order overlap.
If a spectral channel is configured to observe a wavelength $\lambda$ in order $m$, the overlap of diffraction orders would cause the two ``conjugate'' wavelengths $\lambda_\pm$,
\begin{equation}
	\lambda_\pm=\frac{m}{m\pm1}\,\lambda,
\end{equation}
to be imaged at the same location.
In practice, we must ensure that the half-bandwidth of the filter used for observing at wavelength $\lambda$, within which the transmission stays above the maximum acceptable threshold for order suppression (e.g., $0.1\%$), is smaller than
\begin{equation}
	|\lambda-\lambda_+|=\frac{\lambda}{m+1}.
\end{equation}

The ViSP wavelength range can in principle be covered with a set of 18 custom interference filters with a typical \rev{five-cavity} design.
However, many COTS filters are available, and a set of 21 filters was selected from the AVR-Semrock catalog.
Some of these filters have blue or red ``leaks'' that fall within the spectral range of sensitivity of the Andor Zyla 5.5 detectors, and must therefore be used in conjunction with pre-filters that block those leaks.
The list of pre-filters and order-sorting filters available at the time of writing are given in Tables~\ref{tab:prefilters} and~\ref{tab:orderfilters}.

The filters have a clear aperture of $31.5\times63~\mathrm{mm}^2$.
The vertical size of the filters is large enough to capture the full 2-arcmin height of the FOV in anticipation of a possible future upgrade to a camera with $4\mathrm{k}\times4\mathrm{k}$ pixels.
For the same reason, the filters are sized horizontally to capture a spectral range around the observed wavelength wide enough to fill a detector with $2\mathrm{k}$ spectral pixels per beam \rev{(i.e., about $4\mathrm{k}$ spectral pixels total)}.

\renewcommand{\arraystretch}{1.25}
\begin{table}[tbp]
	\begin{tabular}{lrl}
		\hline
		\textbf{AVR Catalog \#} & \textbf{Quantity} & \textbf{Mitigation} \\
		\hline
		\parbox[t]{1.25in}{AVR-0265939\\ 568~nm EdgeBasic LWP} & 1 & blue leak of FF01-775/46 below 550~nm \\
		\parbox[t]{1.25in}{AVR-0265940\\ 532~nm EdgeBasic SWP} & 1 & red leak of FF01-406/15 above 550~nm \\
		\parbox[t]{1.25in}{AVR-0265941\\ 785~nm EdgeBasic SWP} & 3 & red leak of various filters \\
		\hline
	\end{tabular}
	\caption{Blocking pre-filters provided with the ViSP instrument to address blue and red leaks of various order-sorting filters (see Table~\ref{tab:orderfilters}).}
	\label{tab:prefilters}
\end{table}
\renewcommand{\arraystretch}{1}

\renewcommand{\arraystretch}{1.25}
\begin{table}[tbp]
	\begin{tabular}{lrl}
		\hline
		\textbf{AVR Catalog \#} & \textbf{Quantity} & \textbf{Notes} \\
		\hline
		AVR-026542 380/14 BP & 1 & needs 785 SWP \\
		AVR-026543 392/18 BP & 2 & needs 785 SWP \\
		AVR-026544 406/15 BP & 1 & needs 532 SWP \\
		AVR-026545 420/10 BP & 1 & \\
		AVR-026546 433/24 BP & 1 & \\
		AVR-026547 448/20 BP & 1 & needs 785 SWP \\
		AVR-026548 460/14 BP & 1 & needs 785 SWP \\
		AVR-026549 482/25 BP & 1 & needs 785 SWP \\
		AVR-026551 500/24 BP & 1 & needs 785 SWP \\
		AVR-026552 520/15 BP & 1 & needs 785 SWP \\
		AVR-026553 539/30 BP & 1 & \\
		AVR-026554 565/24 BP & 1 & needs 785 SWP \\
		AVR-026555 593/46 BP & 1 & needs 785 SWP \\
		AVR-026556 623/32 BP & 2 & \\
		AVR-026557 655/40 BP & 1 & needs 785 SWP \\
		AVR-026558 697/58 BP & 1 & needs 785 SWP \\
		AVR-026559 732/68 BP & 1 & \\
		AVR-026560 775/46 BP & 1 & needs 568 LWP \\
		AVR-026561 819/44 BP & 1 & \\
		AVR-026562 857/30 BP & 2 & \\
		AVR-026563 889/42 BP & 1 & \\
		\hline
	\end{tabular}
	\caption{Order-sorting filters.
	The catalog number includes a product code followed by the nominal central wavelength and bandpass.}
	\label{tab:orderfilters}
\end{table}
\renewcommand{\arraystretch}{1}

\subsubsection{Polarization Analyzer}\label{sec:analyzer}

The polarizing beam splitter (PBS) and the polarization modulator \citep[not part of the ViSP design effort; see][]{2020JATIS...6c8001H} are the optical elements that make ViSP a polarimeter.
The purpose of the PBS is to analyze the polarization properties of the solar spectrum observed with ViSP.
A discussion on polarimetry and polarization measurement techniques is out of scope for this article.
We refer the reader to \cite{2003isp..book.....D} and \cite{2005aspo.book.....T} for excellent reference works on the subject of polarimetry.

ViSP is a traditional dual-beam polarimeter using a temporal modulation scheme.
This is a common design in which fluctuations of the intensity signal as a result of spurious sources (e.g., atmospheric seeing) are encoded identically in the two orthogonally polarized beams (apart from a possible scaling factor between the two beams), and \rev{they} can therefore be eliminated from the polarization measurement via beam subtraction.
In practice the PBS will not produce exactly orthogonal polarization states in the two beams, but this only impacts the overall polarization efficiency of the instrument and does not impair the ability to perform dual-beam polarimetry \citep[see][for the definition of polarization efficiency]{2000ApOpt..39.1637D}.
A PBS with a $20{:}1$ contrast for both transmitted and reflected beams has an acceptable analyzer efficiency of $\gtrsim90\%$.

\begin{figure}[tbp]
	\centering
	\includegraphics[width=0.75\textwidth]{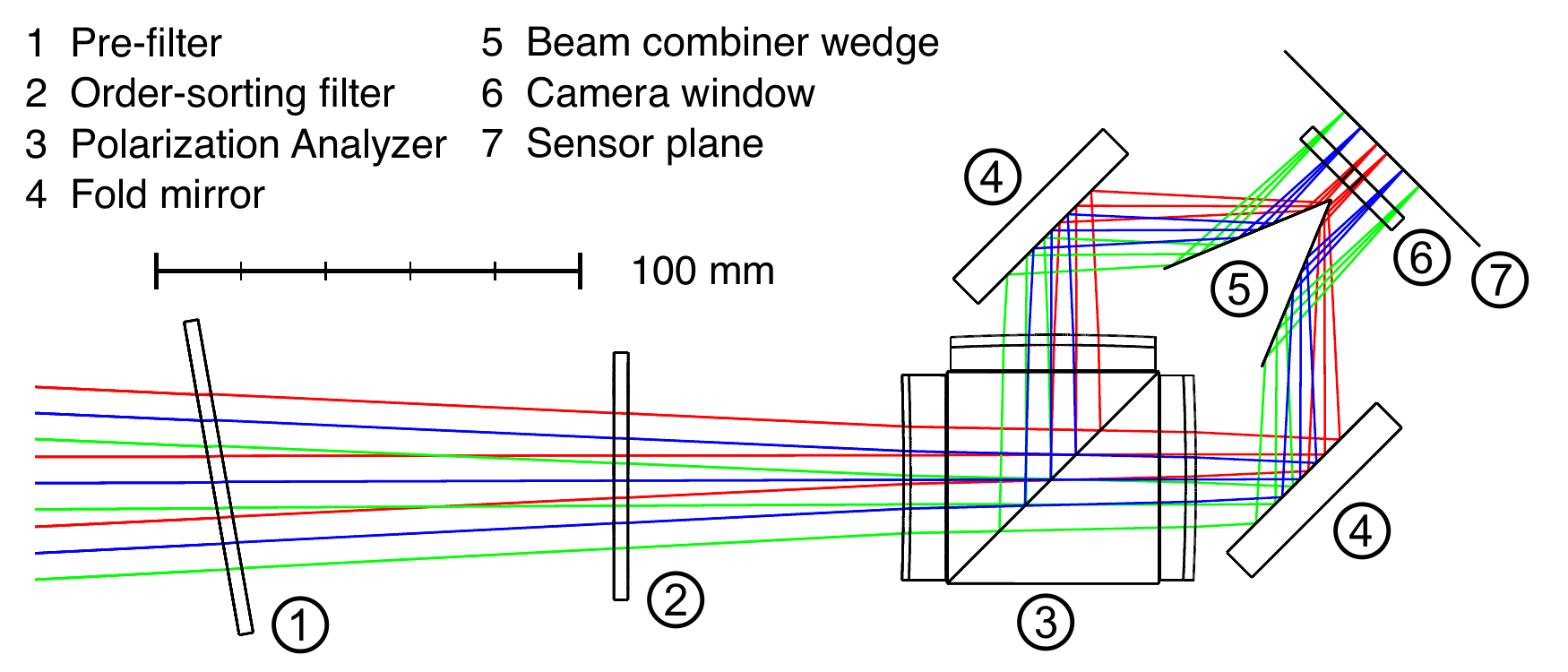}
	\caption{Optical layout of the aft optics assembly.
	The assembly consists \rev{of}, from \textit{bottom-left} to \textit{top-right}, the blocking pre-filter, the order-sorting filter, the beam splitter and relay optics assembly, the beam combiner mirrors and wedge, and the camera entrance window and detector plane.}
	\label{fig:aft_optics}
\end{figure}

The PBS is located close to the camera detector.
Any other location of the PBS along the spectrograph's optical path would have implied a significant increase of the size of optics and/or of the PBS itself.
We note that every arm must be capable of observing any wavelength in the ViSP spectral range in order to satisfy the requirement of spectral versatility.
Consequently, the PBS was replicated identically for each arm.
Like the order-sorting filters, the PBS is sized to accept the full 2~arcmin height of the ViSP FOV.
The PBS is oriented to split \rev{the beam} in the spectral direction in order to maximize the spatial extent of the FOV on the camera.
After splitting, the beams are \rev{recombined} to be imaged side-by-side on a single detector.
Figure~\ref{fig:aft_optics} shows a schematic view of the beam-splitter system.

The range of incidence angles in the spectral dimension at the entrance of the PBS is quite large because of the relatively low $f/\#$ of \rev{about} 8 for the camera lenses, which presents a challenge for the design of a dielectric coating with a contrast better than $20{:}1$ over the entire ViSP spectral range.
It is the largest for \rev{Arm}~1, and amounts to a required total angular acceptance range of approximately $\pm3.8^\circ$, which is too large even when using a high-index glass for the PBS cube.
The possibility of adopting a cube PBS utilizing a wire-grid polarizer at the hypotenuse was considered but discarded because of the large wavefront error that these devices typically have in the reflected beam.
Therefore, a system of relay optics was added around the PBS with the purpose of slowing down the beam inside the PBS.
These relay lenses were optimized simultaneously with the camera lenses (\rev{Section}~\ref{sec:cameras}) for image quality and low angle-of-incidence on the beamsplitter hypotenuse, while constraining overall spectrograph magnification for each channel.
Singlet plano-concave and plano-convex lenses at the PBS entrance and exit faces, respectively, reduce the required angular acceptance range of the PBS to approximately $\pm2.9^\circ$.

The PBS and the relay optics are fabricated from Ohara S-LAH65V, a lanthanum glass selected for its high index of $1.8$ and good transmission down to 380~nm.
The relay optics are bonded to the PBS.
The contrast ratio is above $20{:}1$ at all wavelengths for the transmitted beam.
The reflected beam has higher contrast (around $160{:}1$) over most of the wavelength range, but \rev{the contrast} drops to about $20{:}1$ below about 400~nm.
\rev{The PBS contrast drops quickly at wavelengths above the ViSP spectral range.
This limits the instrument to wavelengths below 900~nm at least for polarimetry, although the instrument optical performance is good up to the IR cut-off wavelength of the silicon detector.}

Two reflections off flat surfaces are used to \rev{recombine} the beams on the detector (see \rev{Figure}~\ref{fig:aft_optics}).
The first reflection is at $45^\circ$ and uses a COTS mirror.
The second reflection is on a custom wedge-shaped substrate at a shallow angle of $67.5^\circ$.
The coating on this substrate is optimized for reflection at this angle over the ViSP wavelength range.
The spectra from the two beams mirror each other in the spectral direction, with the longer wavelength towards the center of the detector.
The beam-combiner wedge is designed to lie as close as possible to the entrance window of the camera, by protruding into its aperture.
This allows us to maximize the spectral bandwidth by minimizing the ``blind'' zone between the beams in front of the wedge of the beam combiner.

\section{Modes of Operation}\label{sec:modes}

The ViSP instrument has two modes of operations\rev{:} \emph{polarimetric} and \emph{intensity}.
We discuss the differences between these modes below.
Both modes of operation support rastering a region of the FOV by moving the slit as well as sit-and-stare operation, in which case the slit is stationary and spatial information is only recorded along the slit.

\subsection{Polarimetric \rev{Mode}}

\rev{The} \emph{polarimetric} mode is the standard mode of operation for ViSP.
It uses a highly efficient polychromatic modulator \rev{\citep{2010ApOpt..49.3580T}} that was designed specifically for the ViSP spectral range to encode the Stokes signals from the observed region into temporally modulated \rev{intensity-signal} sequences \citep{2020JATIS...6c8001H}.
Dual-beam polarization analysis results in two intensity signal sequences that contain nearly orthogonal polarization information that are acquired simultaneously on a single detector.
The modulated intensity signals are then demodulated into Stokes vectors, which must subsequently be combined in order to create an image of Stokes vectors with spectral information in one dimension and spatial information in the other.
The modulator spins continuously and the slit is stepped across the FOV to acquire modulation sequences during a polarimetric observation, thus building up a map of a region.
The slit remains stationary between steps until the full modulation sequence with the desired total integration time has been completed.
The typical number of modulation states per modulation cycle (corresponding to a half rotation of the modulator) is \rev{ten}, \rev{although} configurations with as few as \rev{five} states are possible, at the cost of reduced polarization efficiency.
Typical integration times for polarimetric mode may range from about \rev{one second} for on-disk targets and highly efficient spectrograph configurations, to \rev{up to one minute} for dimmer targets off the limb, such as quiescent prominences.
Many modulation cycles are co-added during an integration.
A lower limit to the integration time is imposed by the time for the modulator to complete a half rotation.
The maximum rotation rate of the ViSP modulator is 5~Hz (300~rpm), giving a theoretical minimum integration time of 0.1~\rev{second}.
This can be used to perform fast time series of a region, but with correspondingly low signal-to-noise ratio (SNR), e.g, for investigations of wave phenomena.
However, the duty cycle of such fast polarimetric operations is low ($<30\%$) except in a sit-and-stare observation where the slit is stationary, because of the 0.2~s necessary to step the slit and settle to a new position, and the polarization efficiency suffers from the \rev{small} number of states that sample the continuously spinning modulator.

\subsection{Intensity \rev{Mode}}

\rev{The} \emph{intensity} mode differs from \rev{the} polarimetric mode in that the slit is scanned at a constant rate of speed across the FOV, rather than being stepped from one position to another.
Polarization measurements are not feasible in this mode because the modulation states would correspond to different areas of the solar image and hence produce an inconsistent data set.
Therefore in this mode the modulator is stopped.
An intensity-only measurement that is, to first order, not contaminated with polarization is recovered by summing the two oppositely polarized beams.

The continuous motion of the slit implies that there is some smearing of the image during the exposure in the direction perpendicular to the slit, but this effect is small if the scan rate does not exceed the width of the slit times the exposure time.
The benefit of this mode is that it permits very fast rastering of the FOV.
For example, a typical scan speed would be around \rev{$1.5~\mathrm{arcsec}\,\mathrm{s}^{-1}$} when using the 650~nm slit, so that the entire FOV of 2~arcmin can be rastered in about 80~\rev{seconds}.
Higher scan speeds are possible at the expense of spatial resolution by choosing a faster scan rate.
A wider slit can also be used, which results in a reduction of both spatial and spectral resolution.
However, a wider slit may also require a camera duty cycle smaller than \rev{unity}, thus reducing the amount of image smearing.

A sequence of maps can be acquired quickly, enabling studies of dynamic events and waves that require high-cadence \rev{time-series} observations.
Some time is lost while the slit returns back to the start of the map, and ViSP prepares to start the next map.
For instance, using the above example of a full FOV raster, the map cadence will be about 100~\rev{seconds}.
The power of intensity mode is evident if we compare this to polarimetric mode: a similar 2~arcmin map at the fastest possible rate in polarimetric mode would result in a map cadence of nearly 15~\rev{minutes}, and with more typical integration times of several seconds such a map would take two hours.

\section{Software}

\subsection{Instrument Control Software}

The ViSP \rev{\textit{Instrument Control System}} (ICS) follows the DKIST software architecture and integrates tightly into the observatory.
It uses the DKIST \rev{\textit{Common Services Framework}} \citep[CSF\rev{:}][]{2010SPIE.7740E..2RH} that is intended to provide easy integration into the facility.
The CSF is used in most major DKIST software systems and was developed to have a common software infrastructure throughout the facility in order to reduce \rev{software-development} effort and maintenance.
The ICS can be used to operate ViSP stand-alone, but as with all facility instruments, the \rev{instrument-control} GUI is also imported into the \rev{\textit{Observatory Control System}} \citep{2012SPIE.8451E..0JJ}, allowing a telescope operator to control ViSP and other instruments through a single interface \citep{2016AN....337.1064T}.

The ViSP ICS can be used by the operator to move mechanisms while showing positions in real time.
It also provides control \rev{of} rack-mounted \rev{power-distribution} units that can be used to switch power to power supplies, computers, and other components in the electronics racks.

The three ViSP cameras and the polarization modulator are provided by DKIST and commanded by the ViSP ICs but controlled through the \rev{\textit{Camera System Software}} (CSS) and the \rev{\textit{Polarization Modulator Controller}} (PMC).
The CSS and PMC provide common interfaces to DKIST cameras and polarization modulators, and, like the CSF, are intended to simplify \rev{software-development} efforts and maintenance.
The strict synchronization between slit motion, modulator rotation, and camera exposures that is required for ViSP operation is handled by the facility time reference and distribution system \citep{2014SPIE.9152E..0ZF}.

\subsection{Detailed Display and Ancillary Processing Plugins}

ViSP provides Detailed Display plugins for the operator to assess data quality.
For polarimetric-mode science data, the plugin can show the modulation states, demodulated data, or maps of Stokes vector components at a specified wavelength pixel that are built up in real time.
In intensity mode, the plugin shows a map of intensity at a specified wavelength pixel.

Ancillary processing plugins are provided to analyze data from focus and alignment calibration tasks in \rev{real time}
The focus plugin determines the best focus position using the Sobel operator to find the camera position at which the image has the highest gradients.
The alignment calibration consists of two parts.
One plugin finds the telescope boresight by calculating the center-of-mass of a pinhole inserted at the \rev{\textit{Gregorian Optical Station}} (GOS) at the secondary Gregorian focus of the telescope \rev{\citep{2020SoPh..295..172R}}.
Another plugin analyzes observations of a \rev{line-grid} target at the GOS using Hough transforms \citep{Duda1972UseOT} to find camera orientation with respect to the direction of dispersion and the spatial scale of the image at the slit and at the detector.
The output from the two \rev{alignment-calibration} plugins is used to calculate FITS \rev{\textit{World Coordinate System}} metadata that \rev{describe} the \rev{physical coordinates} of the image pixels \citep{2002A&A...395.1061G,2006A&A...446..747G}.

\subsection{Instrument Performance Calculator}

\begin{figure}[tbp]
	\centering
	\includegraphics[width=\textwidth]{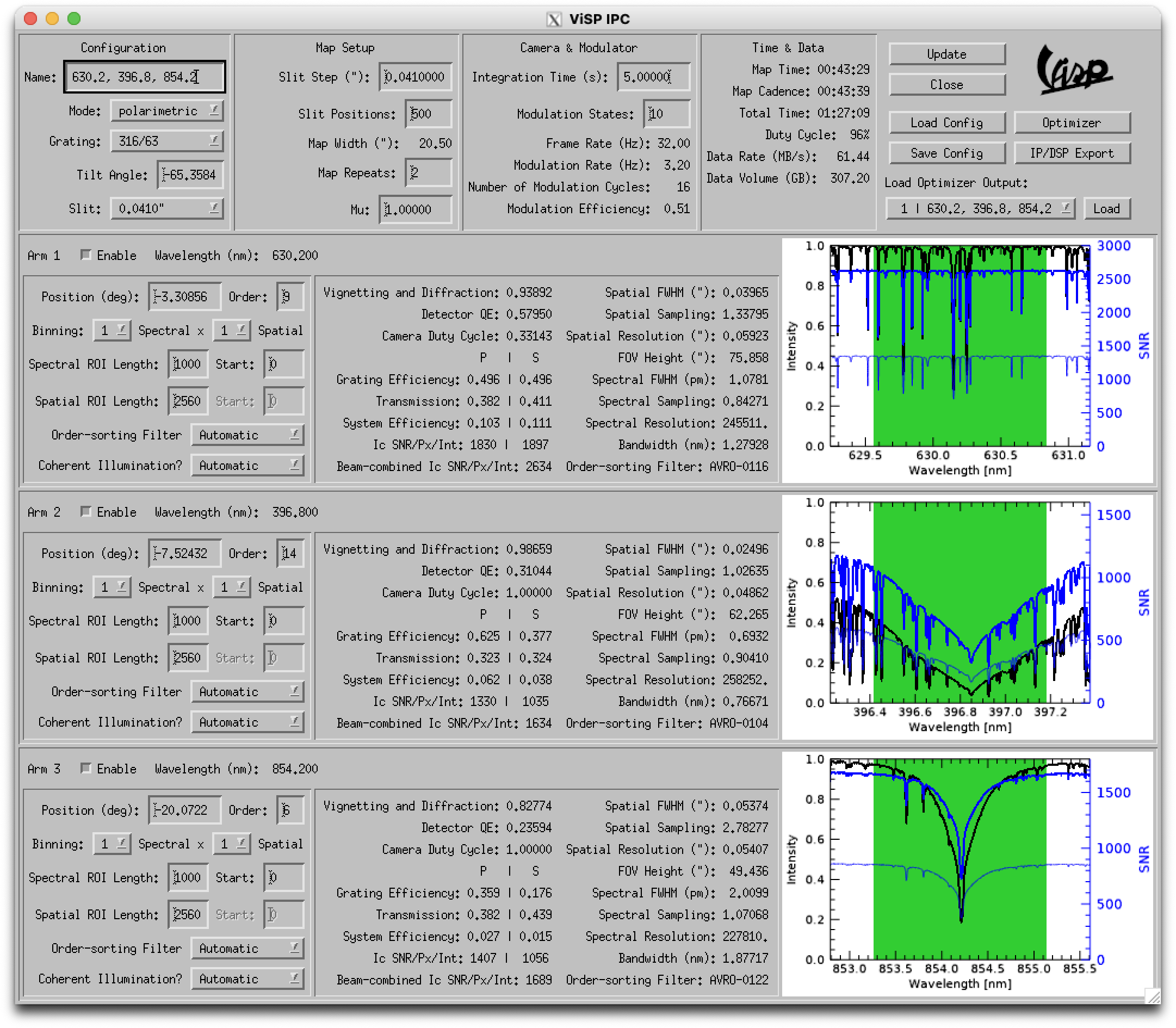}
	\caption{The ViSP IPC GUI interface.
		At the top, the grating, slit, and map parameters can be set, and calculated values such as map duration, cadence, and data volume are shown.
		Parameters and calculated performance figures are displayed for the three arms below.
		In this example, a configuration was optimized with the ICO to observe the \ion{Fe}{1} lines around 630.2~nm, and the \ion{Ca}{2} lines at 396.8 and 854.2~nm.
		The optimizer found a solution with the \ion{Fe}{1} lines in \rev{Arm}~1, and the \ion{Ca}{2} lines in \rev{Arms}~2 and~3.
	A user can interactively change parameters and immediately see the result on the instrument performance.}
	\label{fig:IPC_GUI}
\end{figure}

The ViSP \rev{\textit{Instrument Performance Calculator}} (IPC) is a sophisticated tool to explore the performance of ViSP configurations.
It is intended to enable a user to test the ability of ViSP to address a desired science case.
The IPC calculates \rev{the} expected SNR of the continuum, spatial resolution and field of view, spectral resolution and bandwidth, map size and duration, and data rate and volume for a given configuration of the instrument.
It also shows a spectrum and curves of the SNR as a function of wavelength based on the spectral atlas \rev{of} \cite{1984SoPh...90..205N}.
The IPC uses data for the solar radiation flux at the top of \rev{Haleakal\=a}, and a model of the DKIST and ViSP optical throughput based on measurements of reflectivity and transmission of optics where possible.

The IPC was implemented in the Interactive Data Language\rev{\textsuperscript{\tiny\textregistered}} with a graphical user interface shown in \rev{Figure}~\ref{fig:IPC_GUI}.
The interface allows the user full control over the spectrograph configuration.
It consequently takes a large number of inputs: the desired mode of operation (see \rev{Section}~\ref{sec:modes}), the slit width, the grating (currently only the 316/63 grating is available), the grating tilt angle $\alpha$ (see \rev{Figure}~\ref{fig:layout}), mapping parameters (step size and positions for polarimetric mode, and velocity and scan range for intensity mode), map repeats, solar \rev{$\mu$-angle}, exposure parameters (integration time and modulation states for polarimetric mode, and frame rate for intensity mode), and for each camera, the arm position angle $\delta$, diffraction order, binning, and region-of-interest parameters.
It is also possible to override the automatic selection of the order-sorting filter and the choice of using a calculation under the assumption of coherent and incoherent illumination of the slit aperture.
Instrument configurations generated by the IPC can be saved to file, and, to reduce the risk of mistakes in instrument setup, instrument programs can be exported that can subsequently be loaded by the ICS.

Because of ViSP's ability to tune continuously over the visible solar spectrum in \emph{each} of its spectral channels, determining the most efficient configuration of the spectrograph for a desired set of wavelength regions requires optimization of the spectrograph performance over a large parameter space, consisting of the grating angle, the diffraction order for each of the specified wavelengths, the angular position of each spectral channel, and the pairings of up to three desired wavelengths.
The ViSP IPC is paired with a ViSP \rev{\textit{Instrument Configuration Optimizer}} (ICO) that searches that parameter space for possible spectrograph configurations for a list of lines provided by the user.

The expected performance of ViSP was analyzed using both detailed numerical modeling (e.g., using the Zemax\rev{\textsuperscript{\tiny\textregistered}} \rev{optical-design} package) and general analytic models of the instrument design.
While the numerical modeling was critical to optimize ViSP to its final optical prescription, the general concept for the instrument was originally tested mostly \rev{with} the aid of analytic models.
The ICO is based at its core on the analytic model of the diffraction properties of ViSP \citep{CdW14}, which can predict the throughput of the instrument and its spectral and spatial resolving power for arbitrary configurations of the spectrograph using different slit apertures, grating angles, and angular positions of the camera arms, under both conditions of coherent and incoherent illumination of the slit aperture.
The optimization merit function maximizes the spectrograph efficiency at the desired wavelengths constrained by the ViSP design to only allow configurations that do not lead to mechanical interference among the spectral channels.
The optimizer reads in, and interpolates over, a database of modeled \rev{polarized} efficiencies, which reproduce laboratory measurements \citep{Ca18} with sufficient precision for planning observations and instrument setups.
The output of the ICO can be conveniently imported into the IPC for further analysis.

\section{Example Data}

ViSP acquired on-Sun observations several times during its installation.
Here, we show some of the data from the Science Verification (SV) campaign from \rev{8~May 2021}, of NOAA AR~12822.
For SV, ViSP was set up to observe a band around the well-known \ion{Fe}{1} lines at 630.2~nm in \rev{Arm}~1, the \ion{Ca}{2} H~line at 396.8~nm in \rev{Arm}~2, and the \ion{Ca}{2} line at 854.2~nm.
The configuration is shown in the screenshot of the IPC in \rev{Figure}~\ref{fig:IPC_GUI}.
Both data in polarimetric mode and intensity mode were acquired with the DKIST \textit{Wavefront Correction System} locked most of the time for high-order correction, under reasonable seeing conditions.
These data were processed with the ViSP data calibration pipeline implemented to run in the DKIST Data Center.

\begin{figure}[tbp]
	\centering
	\includegraphics[width=\textwidth]{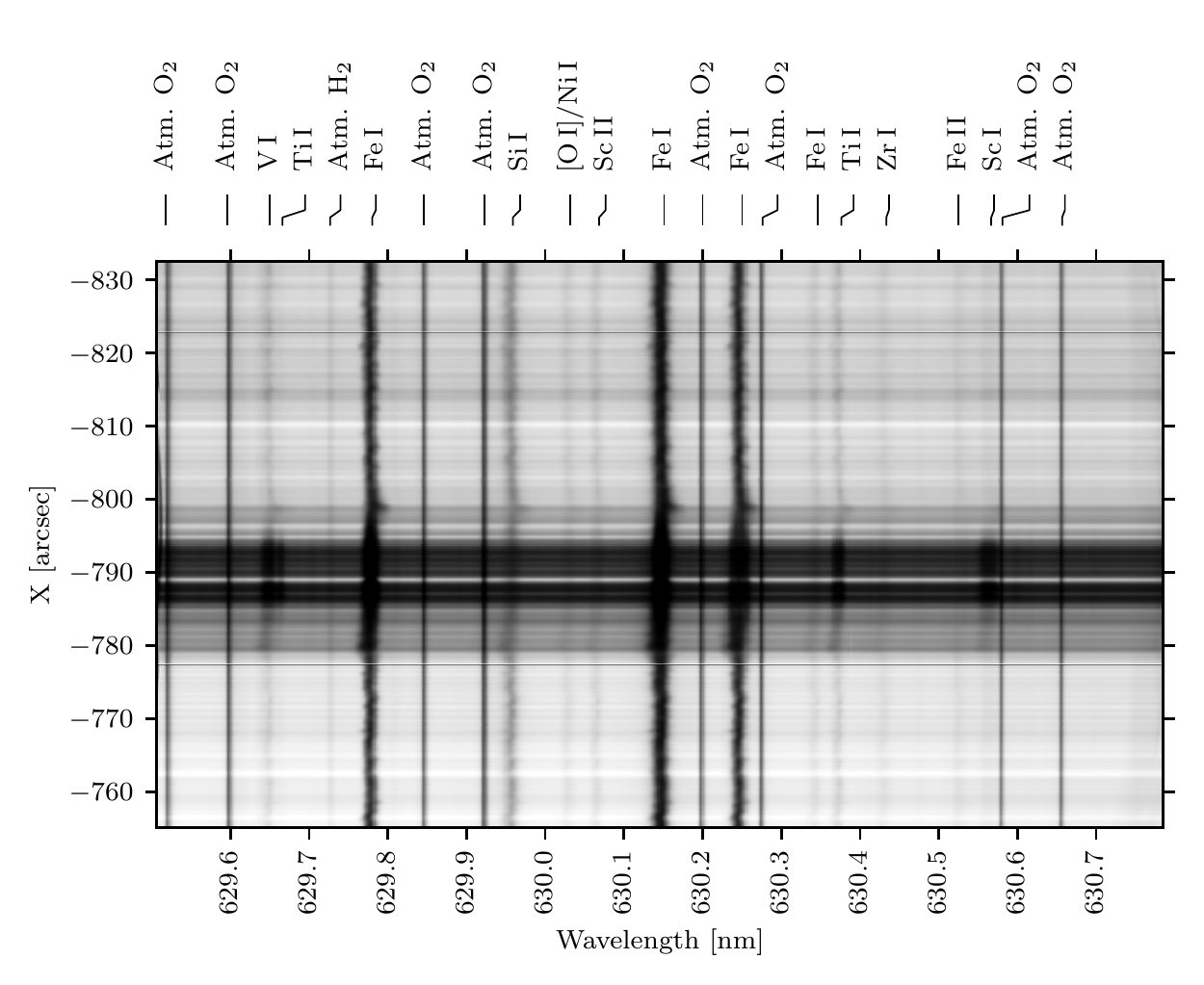}
	\caption{Example \rev{Level-1} data from the ViSP Science Verification campaign, showing Stokes~$I$ over the full spectral bandwidth camera \rev{Arm}~1, with prominent lines identified.
	\rev{The image is scaled by the square of the value to enhance contrast.}}
	\label{fig:level1_full_I}
\end{figure}

\begin{figure}[tbp]
	\centering
	\includegraphics[width=\textwidth]{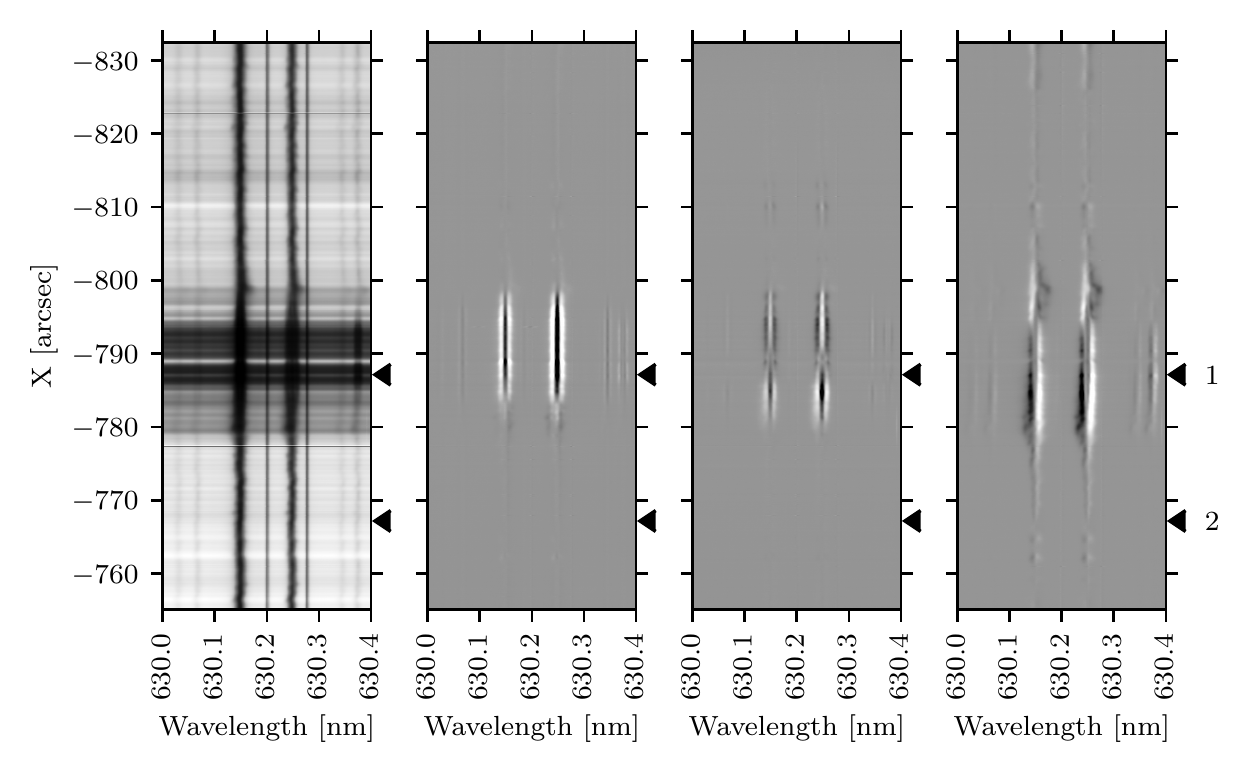}
	\caption{The same example \rev{Level-1} data as in \rev{Figure}~\ref{fig:level1_full_I}, but now showing all Stokes parameters in a spectral region around the \ion{Fe}{1} lines at 630.2~nm.
		\rev{\textit{From left to right}}: Stokes $I$, $Q/I$, $U/I$, and $V/I$.
		\rev{The Stokes-$I$ image is scaled by the square of the value to enhance contrast.}
		The \rev{polarization images} are scaled symmetrically \rev{around zero} and saturated at an amplitude of $20\%$.
	\rev{\textit{Arrows}} indicate the locations of the Stokes profiles shown in \rev{Figure}~\ref{fig:Stokes_630nm}.}
	\label{fig:level1}
\end{figure}

\begin{figure}[tbp]
	\centering
	\includegraphics[width=\textwidth]{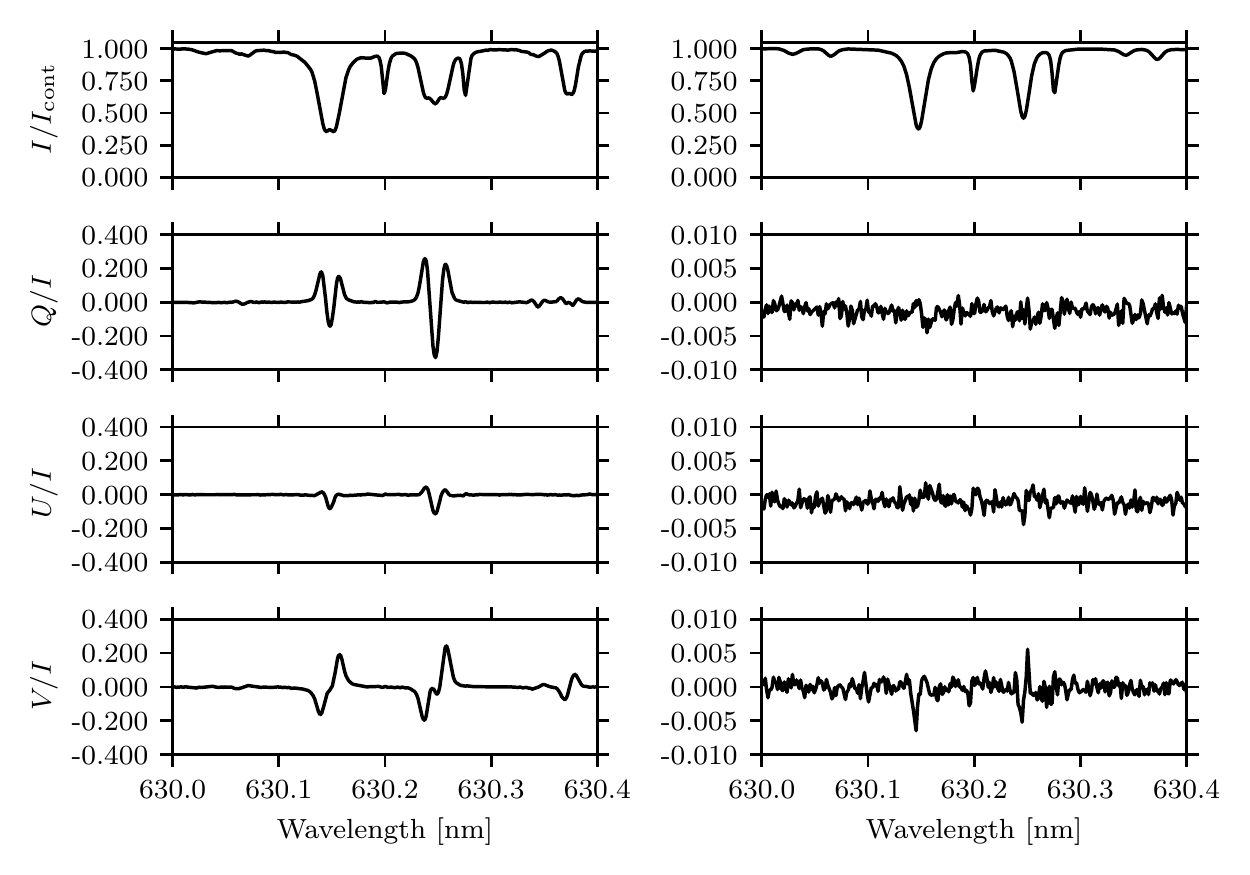}
	\caption{Two examples of full Stokes profiles around the \ion{Fe}{1} 630.2~nm pair, taken in the sunspot (left, arrow~1 in \rev{Figure}~\ref{fig:level1}) and in a relatively quiet area (right, arrow~2 in \rev{Figure}~\ref{fig:level1}).}
	\label{fig:Stokes_630nm}
\end{figure}

Figure~\ref{fig:level1_full_I} shows an example of ViSP \rev{Stokes-$I$} \rev{Level-1} data.
The full field of view of camera \rev{Arm}~1 is shown, with prominent lines identified using the lists of \cite{1966sst..book.....M} and \cite{1995all..book.....K}.
The 1.28-nm wide range covers several telluric $\mathrm{O_2}$ lines that can be used for accurate wavelength referencing, a few photospheric Zeeman-sensitive \ion{Fe}{1} lines including the widely used pair at 630.2~nm, temperature-sensitive \ion{Ti}{1} lines, and the forbidden [\ion{O}{1}] line that has been used for oxygen abundance studies \citep[see, e.g.,][]{2021A&A...653A.141A}.
Figure~\ref{fig:level1} shows the four Stokes parameters from the same data in a wavelength band covering just the \ion{Fe}{1} lines around 630.2~nm.
These data have been corrected for dark current and gain, polarimetrically demodulated and calibrated, have been corrected for spectral skew and curvature, and have the two polarization beams combined.
The FOV contained regions of relatively quiet Sun and a sunspot.
The ViSP slit can be oriented in any direction on the Sun by setting the angle of the coud\'e lab.
In the SV campaign, the lab was oriented to place the slit along the solar \rev{E-W} direction.

Figure~\ref{fig:Stokes_630nm} shows Stokes spectra from these data at the positions indicated by the arrowheads on the right axes in \rev{Figure}~\ref{fig:level1}.
In the quiet region, $Q$ and $U$ are dominated by noise, while $V$ shows a very weak signal.
In the sunspot, we observe the expected broadening and separation of the line profiles into separate components in $I$, and strong polarization signals are present in $Q$, $U$, and $V$.
\rev{These data reached a continuum intensity noise level of about $0.8\times10^{-3}~I_\mathrm{c}$ with an integration time of $1.5~\mathrm{s}$, but required $7.5~\mathrm{s}$ to acquire.
The low camera duty cycle is the result of the combined effect of the short exposure time required to avoid saturation and the limitations on the frame rate imposed by the camera hardware.}

\begin{figure}[tbp]
	\centering
	\includegraphics[width=\textwidth]{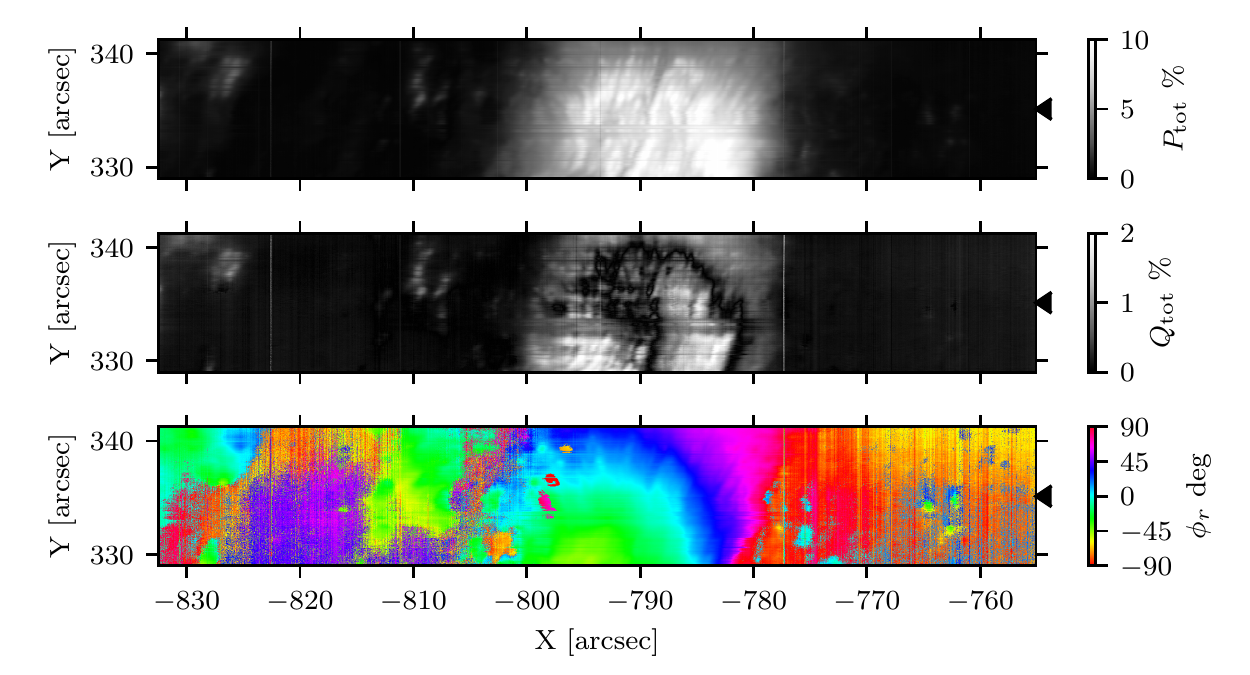}
	\caption{Maps of the total polarization $P_\mathrm{tot}$, net linear polarization $Q_\mathrm{tot}$, and preferred azimuth angle $\phi_r$ from the ViSP Science Verification campaign of NOAA AR~12822, derived from the \ion{Fe}{1} lines at 630.2~nm observed with camera \rev{Arm}~1.
	\rev{\textit{Arrows}} indicate the location of the \rev{Level-1} data shown in \rev{Figure}~\ref{fig:level1}.}
	\label{fig:SV_pol_maps}
\end{figure}

The data shown in \rev{Figure}~\ref{fig:level1} were taken from a scan of 300~steps that covered about 12.3~arcsec on the Sun, using the slit with a width of 0.041~arcsec and a matching \rev{slit-step} distance.
Polarization and field azimuth maps of this region following the technique of \cite{2008ApJ...672.1237L} are shown in \rev{Figure}~\ref{fig:SV_pol_maps}.

\begin{figure}[tbp]
	\centering
	\includegraphics[width=\textwidth]{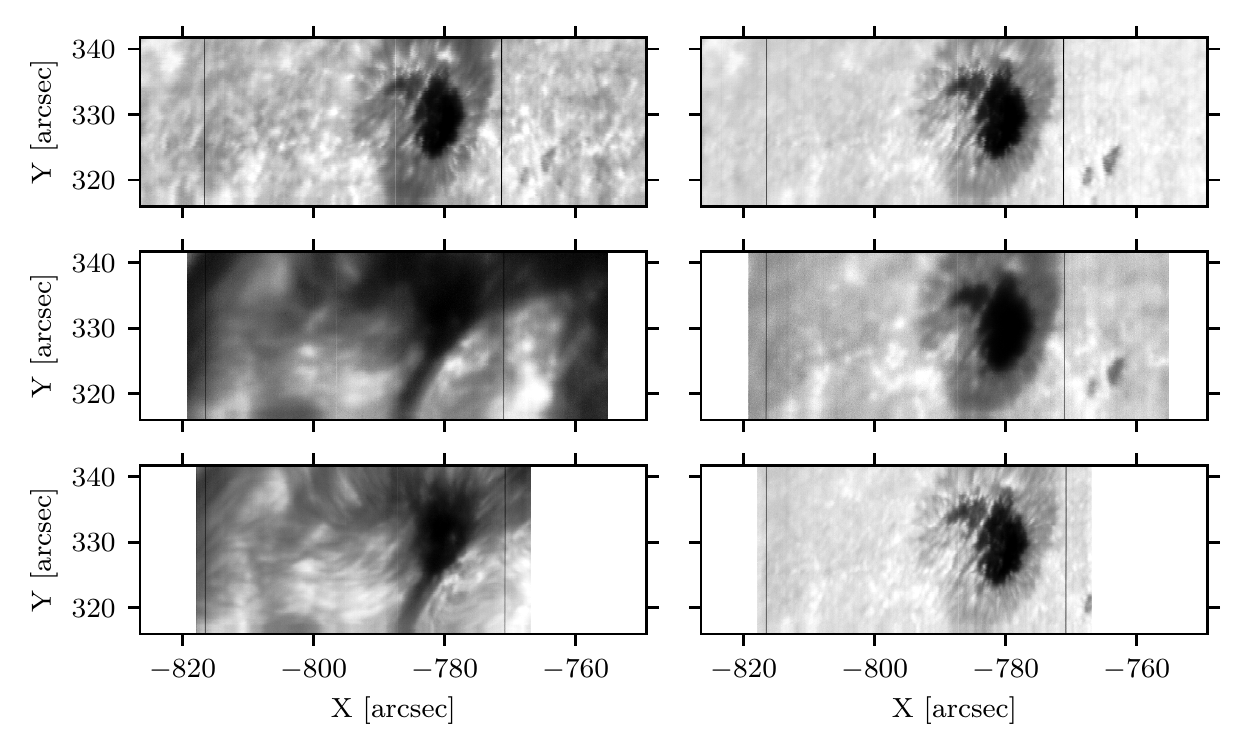}
	\caption{Intensity maps from the ViSP Science Verification campaign of NOAA AR~12822, in the cores (\textit{left}) and the neighboring continuum or far wing (\textit{right}) of the \ion{Fe}{1} 630.25~nm (\textit{top}), \ion{Ca}{2} 396.8~nm (\textit{center}), and \ion{Ca}{2} 854.2~nm (\textit{bottom}) lines, observed with camera arms~1, 2, and~3, respectively.
		The \ion{Ca}{2} 396.8~nm and \ion{Ca}{2} 854.2~nm \rev{line-core} images show the square root of the intensity to enhance contrast in the dark regions.
		\rev{The slit was scanned in the solar N-S direction.}
		We note the different spatial FOV captured by the three arms because of the different spectrograph magnifications in each arm, but the constant hairline separation of 45.2~arcsec.
	\rev{The images are scaled by the square of the value to enhance contrast.}}
	\label{fig:SV_int_maps}
\end{figure}

The same region was also observed in intensity mode, using the same slit as for the polarimetric case.
A series of 37~maps with a larger width of 26~arcsec \rev{was} acquired with a cadence of 34~\rev{seconds}.
Figure~\ref{fig:SV_int_maps} shows sample monochromatic intensity maps from \rev{those} data in the line cores and nearby continuum of the \ion{Fe}{1} 630.25~nm, \ion{Ca}{2} 396.8~nm, and \ion{Ca}{2} 854.2~nm lines.
The continuum images show the photosphere with a sunspot umbra and penumbra, and granulation, as expected.
The \ion{Ca}{2} 396.8~nm is very wide and there is no clean continuum in the 0.77-nm passband (see the atlas spectrum shown in \rev{Figure}~\ref{fig:IPC_GUI}).
The map instead shows the wing of the line where magnetic regions appear with enhanced contrast \citep{2005A&A...437.1069S}.
The \ion{Fe}{1} line core map samples the high photosphere and shows reversed granulation \citep{2012ApJ...757L..17D,2013A&A...556A.127B}.
The two \ion{Ca}{2} \rev{line-core} maps show the chromosphere with its characteristic filamentary structure \citep[see, e.g.,][]{2009A&A...500.1239R}.

\section{Potential Future Upgrades}

The implementation of the ViSP design is inevitably affected by various instrumental effects introduced by departures of several instrument components from their ideal behavior.
This is particularly important with regard to the polarimetric performance of the instrument, because of the dominant role of polarization science at the DKIST.
Other impacts on instrument performance may come from design \rev{descopes} that were required to address cost, schedule, or fabrication risks.
Below, we discuss several potential future upgrades \rev{in no particular order} that would improve the performance and/or capabilities of ViSP.

\subsection{Grating \rev{Polarization}}

As shown in \rev{Section}~\ref{sec:grating}, low-order gratings such as the one adopted for ViSP may introduce significant levels of polarization when $\lambda$ becomes comparable with the grating period $d$ (say, for $\lambda \gtrsim 0.1d$).
The fact that the grating may partially act as a polarization analyzer, before the light reaches the PBS in front of the ViSP detector, poses a risk for the polarimetric performance of the instrument, since it introduces a beam imbalance that reduces the efficiency of dual-beam polarimetry.

In order to see this, we can simply model the grating as a partial polarizer with contrast $0\le p\le 1$, such that the grating is perfectly non-polarizing for $p=0$, and a perfect $Q$-polarizer for $p=1$.
If $\bm{S}=(I,Q,U,V)$ is the Stokes vector of the incident beam on the grating, and any instrumental polarization beyond the grating can be neglected, the two beams reaching the detector will have intensities
\begin{displaymath}
	S^\pm=\frac{1}{2}(1\pm p)(I+Q).
\end{displaymath}
Thus, even in the absence of $Q$ polarization in the incoming Stokes vector, large values of $p$ can have a significant impact on the intensity balance between the two beams at the detector.
In some cases, where high polarimetric sensitivity and temporal resolution are required, it may become critical to reduce the impact of grating polarization.

Grating polarization is mainly caused by a dephasing of the efficiency curve of the TM polarization component (perpendicular to the groove direction) of the diffracted light with respect to the TE component (parallel to the groove direction).
In order to describe and quantify these effects one must employ a vector model of grating diffraction \cite[e.g.,][]{Li99}.
Figure~\ref{fig:grateff} shows an example of such a model output, predicting the presence of spectral regions and diffraction orders with strong polarization.

Strategies to mitigate grating polarization exist, and one that involves a special coating technique of the grating surface \cite[``shadow cast'' coating\rev{:}][]{KM66} has proven to be successful on \rev{small} samples measured in the laboratory \citep{Ca18}.
This type of special coating is being considered for a possible future upgrade of the ViSP grating.

An alternative approach to mitigating the effects of grating polarization is to circumvent the partial $Q$-analysis performed by the grating by transforming the incoming Stokes vector.
A simplistic solution is to introduce two quarter-wave retarders at $45^\circ$ before and after the grating.
The two beams at the detector will then have intensities
\begin{displaymath}
	S^\pm=\frac{1}{2}\left(I-pV\mp\sqrt{1-p^2}\,Q\right),
\end{displaymath}
and thus the two beams are perfectly balanced in the absence of $Q$ polarization, as desired.
A more general approach that eliminates the need for super-achromatic $\lambda/4$ retarders is to optimize two stacks of retarders, one of which rotates $Q$ into some combination of $U$ and $V$ prior to the grating, while the other rotates it back to $Q$ after, over the desired wavelength range \citep{HarringtonGratingPol}.

\subsection{Filter Jukebox}

The original design of ViSP included an automated mechanism (a ``jukebox'') for the selection of the \rev{order-sorting} filters.
Because all three spectral channels have the ability to tune anywhere within the full spectral range of the instrument, this automated approach to filter selection implied the need to have three identical sets of filters.
Because of cost and fabrication risks, the filter jukebox feature was descoped, and only one full set of COTS filters was acquired.

This has an impact on the use of ViSP to observe spectrally close diagnostics that cannot be captured within the typical bandwidth of a single channel (e.g., the simultaneous observations of the \ion{Ca}{2} lines at 849.8 and 854.2~nm).
In the as-built ViSP\rev{,} an operator must enter the coud\'e lab to manually change filters for a change in the ViSP configuration, which limits the ability of the instrument to be rapidly reconfigured for targets of opportunity such as flares, or to execute multiple observing programs with different selections of lines in quick succession.

Some of the COTS filters used have either blue or red leaks that need to be blocked to prevent signal contamination (see \rev{Section}~\ref{sec:filters}).
This has required the addition of \rev{an} edge blocking filters to be used in combination with the ``leaky'' bandpass filters.
A future implementation of the \rev{filter-jukebox} mechanism would therefore require the procurement of three full sets of custom bandpass filters that do not have leaks in the sensitivity range of the ViSP cameras.

We note that three spare filters corresponding to the Science Verification configuration were acquired late in the project, making the above example case of observing two of the \ion{Ca}{2} IR-triplet lines possible (see Table~\ref{tab:orderfilters}).

\subsection{Cameras}

The Andor Zyla 5.5 camera used in ViSP was relatively new at the time that the ViSP design was developed, but more powerful cameras have become commonly available.
The ViSP design could not be adapted to use the Andor Balor camera that is used by other DKIST instruments because its larger pixel size presented too much cost and schedule risk (see \rev{Section}~\ref{sec:cameras}).
However, science-grade cameras with $4\mathrm{k}\times4\mathrm{k}$ pixels with a size of about $6.5~\mathrm{\mu m}$ are now available.
Such a camera could replace the Andor Zyla 5.5 without the need to replace optics (except for the \rev{beam-combiner} wedge), and if the camera form factor is similar, would require limited mechanical \rev{redesign} work.
Modern cameras are also capable of running at higher frame rates, which has the benefit of reducing polarimetric crosstalk \citep{2012ApJ...757...45C}, and would also improve the SNR in spectrograph configurations where a duty cycle \rev{less than $100\%$} must be used to avoid saturation.

\section{Conclusions}

ViSP is one of the four spectro-polarimeters available to the DKIST users, and \rev{it is} currently the only wavelength-versatile instrument capable of tuning on any spectral region of the visible and very-near IR spectrum of the Sun (380 to 900~nm).

Its three spectral channels with automated positioning, a broad-dispersion diffraction grating, and a practically fringe-\rev{free} polychromatic modulator enable users of ViSP to access a virtually infinite number of combinations of spectral lines for the magnetic and plasma diagnostics of the solar atmosphere.
Thanks to its high throughput, ViSP can deliver a polarimetric SNR of 1000 with integration times \rev{less than five seconds} for \rev{on-disk} targets around the peak of the \rev{quantum-efficiency} curve of the Andor Zyla detectors, at the maximum spatial resolution of 0.028~arcsec accessible with the narrowest slit.
The library of five slit apertures allows the instrument to be optimally adapted to the science case at hand, also granting access to dim targets such as prominences and coronal loops.

ViSP has concluded its integration, testing, and commissioning phase with the Science Verification campaign of May 2021, and it is ready to begin science observations for the first year of commissioning of the DKIST facility \rev{that started} in November 2021.

\begin{acks}
	The research reported herein is based in part on data collected with the \textit{Daniel K. Inouye Solar Telescope} (DKIST), a facility of the National Solar Observatory (NSO). NSO is managed by the Association of Universities for Research in Astronomy, Inc., and is funded by the National Science Foundation. Any opinions, ﬁndings and conclusions or recommendations expressed in this publication are those of the authors and do not necessarily reﬂect the views of the National Science Foundation or the Association of Universities for Research in Astronomy, Inc. DKIST is located on land of spiritual and cultural signiﬁcance to Native Hawaiian people. The use of this important site to further scientiﬁc knowledge is done so with appreciation and respect.
	R.~French provided solar coordinates for the SV maps.
	Many people contributed to the ViSP project.
	These include A.~Beard, C.~Berst, G.L.~Card, K.~Cummings, N.~Ela, D.F.~Elmore, A.~Ferayorni, D.~Gallagher, B.~Goodrich, R.~Graves, D.~Harrington, P.~Huang, J.~Hubbard, B.~Larson, M.~Liang, R.~Lull, P.G.~Nelson, P.H.H.~Oakley, P.~Sekulic, H.~Socas-Navarro, K.V.~Streander, S.~Sueoka, R.~Summers, L.~Sutherland, and A.~Watt, amongst others.
	The following acknowledgements were compiled using the Astronomy Acknowledgement Generator (\url{https://astrofrog.github.io/acknowledgment-generator/}).
	This research has made use of NASA's Astrophysics Data System,
	\rev{\textsf{NumPy}} \citep{2011CSE....13b..22V},
	\rev{\textsf{matplotlib}}, a Python library for publication quality graphics \citep{2007CSE.....9...90H},
	\rev{\textsf{Astropy}}, a community-developed core Python package for Astronomy \citep{2018AJ....156..123A, 2013A&A...558A..33A},
	and the \rev{\textsf{IPython}} package \citep{2007CSE.....9c..21P}.
\end{acks}

\begin{authorcontribution}
	Investigation: R.~Casini, A.G.~de Wijn, A.~Carlile, A.R.~Lecinski, S.~Sewell, P.~Zmarzly, C.~Beck, F.~W\"oger;
	Writing - original draft preparation: R.~Casini, A.G.~de Wijn;
	Writing - review and editing: C.~Beck, S.~Sewell, A.D.~Eigenbrot;
	Funding acquisition: R.~Casini, M.~Kn\"olker;
	Software: A.D.~Eigenbrot, A.G.~de Wijn, R.~Casini.
\end{authorcontribution}

\begin{conflict}
	The authors declare that they have no conflicts of interest.
\end{conflict}

\begin{fundinginformation}
	This material is based upon work supported by the National Center for Atmospheric Research, which is a major facility sponsored by the National Science Foundation under Cooperative Agreement No. 1852977.
	Open Access funding is provided by NSO. The National Solar Observatory (NSO) is operated by the Association of Universities for Research in Astronomy, Inc. (AURA), under cooperative agreement with the National Science Foundation.
\end{fundinginformation}

\begin{dataavailability}
	Data sharing is not applicable to this article as no datasets were generated or \rev{analyzed} during the current study.
\end{dataavailability}

\begin{comments}
	This version of the article has been accepted for publication after peer review but is not the Version of Record and does not reflect post-acceptance improvements, or any corrections. The Version of Record is available online at \url{https://dx.doi.org/10.1007/s11207-022-01954-1}
\end{comments}

\bibliographystyle{spr-mp-sola}
\bibliography{main}

\begin{thebibliography}{68}
\ifx\bisbn     \undefined \def\bisbn  #1{ISBN #1}\fi
\ifx\binits    \undefined \def\binits#1{#1}\fi
\ifx\bauthor   \undefined \def\bauthor#1{#1}\fi
\ifx\batitle   \undefined \def\batitle#1{#1}\fi
\ifx\bjtitle   \undefined \def\bjtitle#1{\textit{#1}}\fi
\ifx\bvolume   \undefined \def\bvolume#1{\textbf{#1}}\fi
\ifx\byear     \undefined \def\byear#1{#1}\fi
\ifx\bissue    \undefined \def\bissue#1{#1}\fi
\ifx\bfpage    \undefined \def\bfpage#1{#1}\fi
\ifx\blpage    \undefined \def\blpage #1{#1}\fi
\ifx\burl      \undefined \def\burl#1{#1}\fi
\ifx\href      \undefined \def\href#1#2{#2}\fi
\ifx\betal     \undefined \def\betal{et al.}\fi
\ifx\bctitle   \undefined \def\bctitle#1{#1}\fi
\ifx\beditor   \undefined \def\beditor#1{#1}\fi
\ifx\bbtitle   \undefined \def\bbtitle#1{\textit{#1}}\fi
\ifx\bedition  \undefined \def\bedition#1{#1}\fi
\ifx\bseriesno \undefined \def\bseriesno#1{\textbf{#1}}\fi
\ifx\blocation \undefined \def\blocation#1{#1}\fi
\ifx\bsertitle \undefined \def\bsertitle#1{\textit{#1}}\fi
\ifx\bsnm      \undefined \def\bsnm#1{#1}\fi
\ifx\bsuffix   \undefined \def\bsuffix#1{#1}\fi
\ifx\bparticle \undefined \def\bparticle#1{#1}\fi
\ifx\barticle  \undefined \def\barticle#1{}\fi
\ifx\binstitute  \undefined \def\binstitute#1{#1}\fi
\ifx\bpublisher  \undefined \def\bpublisher#1{#1}\fi
\ifx\doiurl    \undefined \def\doiurl#1{\href{#1}{DOI}}\fi
\makeatletter
\def\safeHref#1#2#3{\in@{http}{#2}\ifin@\href{#2}{#3}\else\href{#1#2}{#3}\fi}
\makeatother
\ifx\adsurl    \undefined
  \def\adsurl#1{\safeHref{https://ui.adsabs.harvard.edu/abs/}{#1}{ADS}}\fi
\ifx\arxivurl  \undefined
  \def\arxivurl#1{\safeHref{http://arxiv.org/abs/}{#1}{arXiv}}\fi
\ifx\botherref \undefined \def\botherref#1{}\fi
\ifx\url       \undefined \def\url#1{#1}\fi
\ifx\bchapter  \undefined \def\bchapter#1{}\fi
\ifx\bbook     \undefined \def\bbook#1{}\fi
\ifx\bcomment  \undefined \def\bcomment#1{#1}\fi
\ifx\oauthor   \undefined \def\oauthor#1{#1}\fi
\ifx\citeauthoryear \undefined\def \citeauthoryear#1{#1}\fi
\def\endbibitem {}
\ifx\bconflocation  \undefined \def\bconflocation#1{#1} \fi

\bibitem[\protect\citeauthoryear{{Asplund}, {Amarsi}, and
  {Grevesse}}{2021}]{2021A&A...653A.141A}
\begin{barticle}
\bauthor{\bsnm{{Asplund}}, \binits{M.}},
\bauthor{\bsnm{{Amarsi}}, \binits{A.M.}},
\bauthor{\bsnm{{Grevesse}}, \binits{N.}}:
\byear{2021},
\batitle{{The chemical make-up of the Sun: A 2020 vision}}.
\bjtitle{\aap}
\bvolume{653},
\bfpage{A141}.
\doiurl{https://doi.org/10.1051/0004-6361/202140445}.
\adsurl{2021A&A...653A.141A}.
\end{barticle}
\endbibitem

\bibitem[\protect\citeauthoryear{{Astropy Collaboration}
  et~al.}{2013}]{2013A&A...558A..33A}
\begin{barticle}
\bauthor{\bsnm{{Astropy Collaboration}}},
\bauthor{\bsnm{{Robitaille}}, \binits{T.P.}},
\bauthor{\bsnm{{Tollerud}}, \binits{E.J.}},
\bauthor{\bsnm{{Greenfield}}, \binits{P.}},
\bauthor{\bsnm{{Droettboom}}, \binits{M.}},
\bauthor{\bsnm{{Bray}}, \binits{E.}},
\bauthor{\bsnm{{Aldcroft}}, \binits{T.}},
\bauthor{\bsnm{{Davis}}, \binits{M.}},
\bauthor{\bsnm{{Ginsburg}}, \binits{A.}},
\bauthor{\bsnm{{Price-Whelan}}, \binits{A.M.}},
\bauthor{\bsnm{{Kerzendorf}}, \binits{W.E.}},
\bauthor{\bsnm{{Conley}}, \binits{A.}},
\bauthor{\bsnm{{Crighton}}, \binits{N.}},
\bauthor{\bsnm{{Barbary}}, \binits{K.}},
\bauthor{\bsnm{{Muna}}, \binits{D.}},
\bauthor{\bsnm{{Ferguson}}, \binits{H.}},
\bauthor{\bsnm{{Grollier}}, \binits{F.}},
\bauthor{\bsnm{{Parikh}}, \binits{M.M.}},
\bauthor{\bsnm{{Nair}}, \binits{P.H.}},
\bauthor{\bsnm{{Unther}}, \binits{H.M.}},
\bauthor{\bsnm{{Deil}}, \binits{C.}},
\bauthor{\bsnm{{Woillez}}, \binits{J.}},
\bauthor{\bsnm{{Conseil}}, \binits{S.}},
\bauthor{\bsnm{{Kramer}}, \binits{R.}},
\bauthor{\bsnm{{Turner}}, \binits{J.E.H.}},
\bauthor{\bsnm{{Singer}}, \binits{L.}},
\bauthor{\bsnm{{Fox}}, \binits{R.}},
\bauthor{\bsnm{{Weaver}}, \binits{B.A.}},
\bauthor{\bsnm{{Zabalza}}, \binits{V.}},
\bauthor{\bsnm{{Edwards}}, \binits{Z.I.}},
\bauthor{\bsnm{{Azalee Bostroem}}, \binits{K.}},
\bauthor{\bsnm{{Burke}}, \binits{D.J.}},
\bauthor{\bsnm{{Casey}}, \binits{A.R.}},
\bauthor{\bsnm{{Crawford}}, \binits{S.M.}},
\bauthor{\bsnm{{Dencheva}}, \binits{N.}},
\bauthor{\bsnm{{Ely}}, \binits{J.}},
\bauthor{\bsnm{{Jenness}}, \binits{T.}},
\bauthor{\bsnm{{Labrie}}, \binits{K.}},
\bauthor{\bsnm{{Lim}}, \binits{P.L.}},
\bauthor{\bsnm{{Pierfederici}}, \binits{F.}},
\bauthor{\bsnm{{Pontzen}}, \binits{A.}},
\bauthor{\bsnm{{Ptak}}, \binits{A.}},
\bauthor{\bsnm{{Refsdal}}, \binits{B.}},
\bauthor{\bsnm{{Servillat}}, \binits{M.}},
\bauthor{\bsnm{{Streicher}}, \binits{O.}}:
\byear{2013},
\batitle{{Astropy: A community Python package for astronomy}}.
\bjtitle{\aap}
\bvolume{558},
\bfpage{A33}.
\doiurl{https://doi.org/10.1051/0004-6361/201322068}.
\adsurl{2013A&A...558A..33A}.
\end{barticle}
\endbibitem

\bibitem[\protect\citeauthoryear{{Astropy Collaboration}
  et~al.}{2018}]{2018AJ....156..123A}
\begin{barticle}
\bauthor{\bsnm{{Astropy Collaboration}}},
\bauthor{\bsnm{{Price-Whelan}}, \binits{A.M.}},
\bauthor{\bsnm{{Sip{\H{o}}cz}}, \binits{B.M.}},
\bauthor{\bsnm{{G{\"u}nther}}, \binits{H.M.}},
\bauthor{\bsnm{{Lim}}, \binits{P.L.}},
\bauthor{\bsnm{{Crawford}}, \binits{S.M.}},
\bauthor{\bsnm{{Conseil}}, \binits{S.}},
\bauthor{\bsnm{{Shupe}}, \binits{D.L.}},
\bauthor{\bsnm{{Craig}}, \binits{M.W.}},
\bauthor{\bsnm{{Dencheva}}, \binits{N.}},
\bauthor{\bsnm{{Ginsburg}}, \binits{A.}},
\bauthor{\bsnm{{Vand erPlas}}, \binits{J.T.}},
\bauthor{\bsnm{{Bradley}}, \binits{L.D.}},
\bauthor{\bsnm{{P{\'e}rez-Su{\'a}rez}}, \binits{D.}},
\bauthor{\bsnm{{de Val-Borro}}, \binits{M.}},
\bauthor{\bsnm{{Aldcroft}}, \binits{T.L.}},
\bauthor{\bsnm{{Cruz}}, \binits{K.L.}},
\bauthor{\bsnm{{Robitaille}}, \binits{T.P.}},
\bauthor{\bsnm{{Tollerud}}, \binits{E.J.}},
\bauthor{\bsnm{{Ardelean}}, \binits{C.}},
\bauthor{\bsnm{{Babej}}, \binits{T.}},
\bauthor{\bsnm{{Bach}}, \binits{Y.P.}},
\bauthor{\bsnm{{Bachetti}}, \binits{M.}},
\bauthor{\bsnm{{Bakanov}}, \binits{A.V.}},
\bauthor{\bsnm{{Bamford}}, \binits{S.P.}},
\bauthor{\bsnm{{Barentsen}}, \binits{G.}},
\bauthor{\bsnm{{Barmby}}, \binits{P.}},
\bauthor{\bsnm{{Baumbach}}, \binits{A.}},
\bauthor{\bsnm{{Berry}}, \binits{K.L.}},
\bauthor{\bsnm{{Biscani}}, \binits{F.}},
\bauthor{\bsnm{{Boquien}}, \binits{M.}},
\bauthor{\bsnm{{Bostroem}}, \binits{K.A.}},
\bauthor{\bsnm{{Bouma}}, \binits{L.G.}},
\bauthor{\bsnm{{Brammer}}, \binits{G.B.}},
\bauthor{\bsnm{{Bray}}, \binits{E.M.}},
\bauthor{\bsnm{{Breytenbach}}, \binits{H.}},
\bauthor{\bsnm{{Buddelmeijer}}, \binits{H.}},
\bauthor{\bsnm{{Burke}}, \binits{D.J.}},
\bauthor{\bsnm{{Calderone}}, \binits{G.}},
\bauthor{\bsnm{{Cano Rodr{\'\i}guez}}, \binits{J.L.}},
\bauthor{\bsnm{{Cara}}, \binits{M.}},
\bauthor{\bsnm{{Cardoso}}, \binits{J.V.M.}},
\bauthor{\bsnm{{Cheedella}}, \binits{S.}},
\bauthor{\bsnm{{Copin}}, \binits{Y.}},
\bauthor{\bsnm{{Corrales}}, \binits{L.}},
\bauthor{\bsnm{{Crichton}}, \binits{D.}},
\bauthor{\bsnm{{D'Avella}}, \binits{D.}},
\bauthor{\bsnm{{Deil}}, \binits{C.}},
\bauthor{\bsnm{{Depagne}}, \binits{{\'E}.}},
\bauthor{\bsnm{{Dietrich}}, \binits{J.P.}},
\bauthor{\bsnm{{Donath}}, \binits{A.}},
\bauthor{\bsnm{{Droettboom}}, \binits{M.}},
\bauthor{\bsnm{{Earl}}, \binits{N.}},
\bauthor{\bsnm{{Erben}}, \binits{T.}},
\bauthor{\bsnm{{Fabbro}}, \binits{S.}},
\bauthor{\bsnm{{Ferreira}}, \binits{L.A.}},
\bauthor{\bsnm{{Finethy}}, \binits{T.}},
\bauthor{\bsnm{{Fox}}, \binits{R.T.}},
\bauthor{\bsnm{{Garrison}}, \binits{L.H.}},
\bauthor{\bsnm{{Gibbons}}, \binits{S.L.J.}},
\bauthor{\bsnm{{Goldstein}}, \binits{D.A.}},
\bauthor{\bsnm{{Gommers}}, \binits{R.}},
\bauthor{\bsnm{{Greco}}, \binits{J.P.}},
\bauthor{\bsnm{{Greenfield}}, \binits{P.}},
\bauthor{\bsnm{{Groener}}, \binits{A.M.}},
\bauthor{\bsnm{{Grollier}}, \binits{F.}},
\bauthor{\bsnm{{Hagen}}, \binits{A.}},
\bauthor{\bsnm{{Hirst}}, \binits{P.}},
\bauthor{\bsnm{{Homeier}}, \binits{D.}},
\bauthor{\bsnm{{Horton}}, \binits{A.J.}},
\bauthor{\bsnm{{Hosseinzadeh}}, \binits{G.}},
\bauthor{\bsnm{{Hu}}, \binits{L.}},
\bauthor{\bsnm{{Hunkeler}}, \binits{J.S.}},
\bauthor{\bsnm{{Ivezi{\'c}}}, \binits{{\v{Z}}.}},
\bauthor{\bsnm{{Jain}}, \binits{A.}},
\bauthor{\bsnm{{Jenness}}, \binits{T.}},
\bauthor{\bsnm{{Kanarek}}, \binits{G.}},
\bauthor{\bsnm{{Kendrew}}, \binits{S.}},
\bauthor{\bsnm{{Kern}}, \binits{N.S.}},
\bauthor{\bsnm{{Kerzendorf}}, \binits{W.E.}},
\bauthor{\bsnm{{Khvalko}}, \binits{A.}},
\bauthor{\bsnm{{King}}, \binits{J.}},
\bauthor{\bsnm{{Kirkby}}, \binits{D.}},
\bauthor{\bsnm{{Kulkarni}}, \binits{A.M.}},
\bauthor{\bsnm{{Kumar}}, \binits{A.}},
\bauthor{\bsnm{{Lee}}, \binits{A.}},
\bauthor{\bsnm{{Lenz}}, \binits{D.}},
\bauthor{\bsnm{{Littlefair}}, \binits{S.P.}},
\bauthor{\bsnm{{Ma}}, \binits{Z.}},
\bauthor{\bsnm{{Macleod}}, \binits{D.M.}},
\bauthor{\bsnm{{Mastropietro}}, \binits{M.}},
\bauthor{\bsnm{{McCully}}, \binits{C.}},
\bauthor{\bsnm{{Montagnac}}, \binits{S.}},
\bauthor{\bsnm{{Morris}}, \binits{B.M.}},
\bauthor{\bsnm{{Mueller}}, \binits{M.}},
\bauthor{\bsnm{{Mumford}}, \binits{S.J.}},
\bauthor{\bsnm{{Muna}}, \binits{D.}},
\bauthor{\bsnm{{Murphy}}, \binits{N.A.}},
\bauthor{\bsnm{{Nelson}}, \binits{S.}},
\bauthor{\bsnm{{Nguyen}}, \binits{G.H.}},
\bauthor{\bsnm{{Ninan}}, \binits{J.P.}},
\bauthor{\bsnm{{N{\"o}the}}, \binits{M.}},
\bauthor{\bsnm{{Ogaz}}, \binits{S.}},
\bauthor{\bsnm{{Oh}}, \binits{S.}},
\bauthor{\bsnm{{Parejko}}, \binits{J.K.}},
\bauthor{\bsnm{{Parley}}, \binits{N.}},
\bauthor{\bsnm{{Pascual}}, \binits{S.}},
\bauthor{\bsnm{{Patil}}, \binits{R.}},
\bauthor{\bsnm{{Patil}}, \binits{A.A.}},
\bauthor{\bsnm{{Plunkett}}, \binits{A.L.}},
\bauthor{\bsnm{{Prochaska}}, \binits{J.X.}},
\bauthor{\bsnm{{Rastogi}}, \binits{T.}},
\bauthor{\bsnm{{Reddy Janga}}, \binits{V.}},
\bauthor{\bsnm{{Sabater}}, \binits{J.}},
\bauthor{\bsnm{{Sakurikar}}, \binits{P.}},
\bauthor{\bsnm{{Seifert}}, \binits{M.}},
\bauthor{\bsnm{{Sherbert}}, \binits{L.E.}},
\bauthor{\bsnm{{Sherwood-Taylor}}, \binits{H.}},
\bauthor{\bsnm{{Shih}}, \binits{A.Y.}},
\bauthor{\bsnm{{Sick}}, \binits{J.}},
\bauthor{\bsnm{{Silbiger}}, \binits{M.T.}},
\bauthor{\bsnm{{Singanamalla}}, \binits{S.}},
\bauthor{\bsnm{{Singer}}, \binits{L.P.}},
\bauthor{\bsnm{{Sladen}}, \binits{P.H.}},
\bauthor{\bsnm{{Sooley}}, \binits{K.A.}},
\bauthor{\bsnm{{Sornarajah}}, \binits{S.}},
\bauthor{\bsnm{{Streicher}}, \binits{O.}},
\bauthor{\bsnm{{Teuben}}, \binits{P.}},
\bauthor{\bsnm{{Thomas}}, \binits{S.W.}},
\bauthor{\bsnm{{Tremblay}}, \binits{G.R.}},
\bauthor{\bsnm{{Turner}}, \binits{J.E.H.}},
\bauthor{\bsnm{{Terr{\'o}n}}, \binits{V.}},
\bauthor{\bsnm{{van Kerkwijk}}, \binits{M.H.}},
\bauthor{\bsnm{{de la Vega}}, \binits{A.}},
\bauthor{\bsnm{{Watkins}}, \binits{L.L.}},
\bauthor{\bsnm{{Weaver}}, \binits{B.A.}},
\bauthor{\bsnm{{Whitmore}}, \binits{J.B.}},
\bauthor{\bsnm{{Woillez}}, \binits{J.}},
\bauthor{\bsnm{{Zabalza}}, \binits{V.}},
\bauthor{\bsnm{{Astropy Contributors}}}:
\byear{2018},
\batitle{{The Astropy Project: Building an Open-science Project and Status of
  the v2.0 Core Package}}.
\bjtitle{\aj}
\bvolume{156},
\bfpage{123}.
\doiurl{https://doi.org/10.3847/1538-3881/aabc4f}.
\adsurl{2018AJ....156..123A}.
\end{barticle}
\endbibitem

\bibitem[\protect\citeauthoryear{{Balthasar} and
  {Demidov}}{2012}]{2012SoPh..280..355B}
\begin{barticle}
\bauthor{\bsnm{{Balthasar}}, \binits{H.}},
\bauthor{\bsnm{{Demidov}}, \binits{M.L.}}:
\byear{2012},
\batitle{{Spectral Inversion of Multiline Full-Disk Observations of Quiet Sun
  Magnetic Fields}}.
\bjtitle{\solphys}
\bvolume{280},
\bfpage{355}.
\doiurl{https://doi.org/10.1007/s11207-012-9981-0}.
\adsurl{2012SoPh..280..355B}.
\end{barticle}
\endbibitem

\bibitem[\protect\citeauthoryear{{Beck}, {Rezaei}, and
  {Puschmann}}{2013}]{2013A&A...556A.127B}
\begin{barticle}
\bauthor{\bsnm{{Beck}}, \binits{C.}},
\bauthor{\bsnm{{Rezaei}}, \binits{R.}},
\bauthor{\bsnm{{Puschmann}}, \binits{K.G.}}:
\byear{2013},
\batitle{{Can spicules be detected at disc centre in broad-band Ca ii H filter
  imaging data?}}
\bjtitle{\aap}
\bvolume{556},
\bfpage{A127}.
\doiurl{https://doi.org/10.1051/0004-6361/201220848}.
\adsurl{2013A&A...556A.127B}.
\end{barticle}
\endbibitem

\bibitem[\protect\citeauthoryear{{Beck} et~al.}{2005}]{2005A&A...437.1159B}
\begin{barticle}
\bauthor{\bsnm{{Beck}}, \binits{C.}},
\bauthor{\bsnm{{Schmidt}}, \binits{W.}},
\bauthor{\bsnm{{Kentischer}}, \binits{T.}},
\bauthor{\bsnm{{Elmore}}, \binits{D.}}:
\byear{2005},
\batitle{{Polarimetric Littrow Spectrograph - instrument calibration and first
  measurements}}.
\bjtitle{\aap}
\bvolume{437},
\bfpage{1159}.
\doiurl{https://doi.org/10.1051/0004-6361:20052662}.
\adsurl{2005A&A...437.1159B}.
\end{barticle}
\endbibitem

\bibitem[\protect\citeauthoryear{{Bianda} et~al.}{2018}]{2018A&A...614A..89B}
\begin{barticle}
\bauthor{\bsnm{{Bianda}}, \binits{M.}},
\bauthor{\bsnm{{Berdyugina}}, \binits{S.}},
\bauthor{\bsnm{{Gisler}}, \binits{D.}},
\bauthor{\bsnm{{Ramelli}}, \binits{R.}},
\bauthor{\bsnm{{Belluzzi}}, \binits{L.}},
\bauthor{\bsnm{{Carlin}}, \binits{E.S.}},
\bauthor{\bsnm{{Stenflo}}, \binits{J.O.}},
\bauthor{\bsnm{{Berkefeld}}, \binits{T.}}:
\byear{2018},
\batitle{{Spatial variations of the Sr I 4607 {\r{A}} scattering polarization
  peak}}.
\bjtitle{\aap}
\bvolume{614},
\bfpage{A89}.
\doiurl{https://doi.org/10.1051/0004-6361/201731887}.
\adsurl{2018A&A...614A..89B}.
\end{barticle}
\endbibitem

\bibitem[\protect\citeauthoryear{{Casini} and {de Wijn}}{2014}]{CdW14}
\begin{barticle}
\bauthor{\bsnm{{Casini}}, \binits{R.}},
\bauthor{\bsnm{{de Wijn}}, \binits{A.G.}}:
\byear{2014},
\batitle{{On the instrument profile of slit spectrographs}}.
\bjtitle{J. Opt. Soc. Am. A}
\bvolume{31},
\bfpage{2002}.
\doiurl{https://doi.org/10.1364/JOSAA.31.002002}.
\adsurl{2014JOSAA..31.2002C}.
\end{barticle}
\endbibitem

\bibitem[\protect\citeauthoryear{{Casini} and {Nelson}}{2014}]{CN14}
\begin{barticle}
\bauthor{\bsnm{{Casini}}, \binits{R.}},
\bauthor{\bsnm{{Nelson}}, \binits{P.G.}}:
\byear{2014},
\batitle{{On the intensity distribution function of blazed reflective
  diffraction gratings}}.
\bjtitle{J. Opt. Soc. Am. A}
\bvolume{31},
\bfpage{2179}.
\doiurl{https://doi.org/10.1364/JOSAA.31.002179}.
\adsurl{2014JOSAA..31.2179C}.
\end{barticle}
\endbibitem

\bibitem[\protect\citeauthoryear{{Casini}, {de Wijn}, and
  {Judge}}{2012}]{2012ApJ...757...45C}
\begin{barticle}
\bauthor{\bsnm{{Casini}}, \binits{R.}},
\bauthor{\bsnm{{de Wijn}}, \binits{A.G.}},
\bauthor{\bsnm{{Judge}}, \binits{P.G.}}:
\byear{2012},
\batitle{{Analysis of Seeing-induced Polarization Cross-talk and Modulation
  Scheme Performance}}.
\bjtitle{\apj}
\bvolume{757},
\bfpage{45}.
\doiurl{https://doi.org/10.1088/0004-637X/757/1/45}.
\adsurl{2012ApJ...757...45C}.
\end{barticle}
\endbibitem

\bibitem[\protect\citeauthoryear{{Casini} et~al.}{2018}]{Ca18}
\begin{barticle}
\bauthor{\bsnm{{Casini}}, \binits{R.}},
\bauthor{\bsnm{{Gallagher}}, \binits{D.}},
\bauthor{\bsnm{{Cordova}}, \binits{A.}},
\bauthor{\bsnm{{Morgan}}, \binits{M.}}:
\byear{2018},
\batitle{{Measured performance of shadow-cast coated gratings for
  spectro-polarimetric applications}}.
\bjtitle{\applopt}
\bvolume{57},
\bfpage{7276}.
\doiurl{https://doi.org/10.1364/AO.57.007276}.
\adsurl{2018ApOpt..57.7276C}.
\end{barticle}
\endbibitem

\bibitem[\protect\citeauthoryear{{Collados} et~al.}{2012}]{2012AN....333..872C}
\begin{barticle}
\bauthor{\bsnm{{Collados}}, \binits{M.}},
\bauthor{\bsnm{{L{\'o}pez}}, \binits{R.}},
\bauthor{\bsnm{{P{\'a}ez}}, \binits{E.}},
\bauthor{\bsnm{{Hern{\'a}ndez}}, \binits{E.}},
\bauthor{\bsnm{{Reyes}}, \binits{M.}},
\bauthor{\bsnm{{Calcines}}, \binits{A.}},
\bauthor{\bsnm{{Ballesteros}}, \binits{E.}},
\bauthor{\bsnm{{D{\'\i}az}}, \binits{J.J.}},
\bauthor{\bsnm{{Denker}}, \binits{C.}},
\bauthor{\bsnm{{Lagg}}, \binits{A.}},
\bauthor{\bsnm{{Schlichenmaier}}, \binits{R.}},
\bauthor{\bsnm{{Schmidt}}, \binits{W.}},
\bauthor{\bsnm{{Solanki}}, \binits{S.K.}},
\bauthor{\bsnm{{Strassmeier}}, \binits{K.G.}},
\bauthor{\bsnm{{von der L{\"u}he}}, \binits{O.}},
\bauthor{\bsnm{{Volkmer}}, \binits{R.}}:
\byear{2012},
\batitle{{GRIS: The GREGOR Infrared Spectrograph}}.
\bjtitle{Astron. Nach.}
\bvolume{333},
\bfpage{872}.
\doiurl{https://doi.org/10.1002/asna.201211738}.
\adsurl{2012AN....333..872C}.
\end{barticle}
\endbibitem

\bibitem[\protect\citeauthoryear{{de Wijn}}{2012}]{2012ApJ...757L..17D}
\begin{barticle}
\bauthor{\bsnm{{de Wijn}}, \binits{A.G.}}:
\byear{2012},
\batitle{{Probable Identification of the On-disk Counterpart of Spicules in
  Hinode Ca II H Observations}}.
\bjtitle{\apjl}
\bvolume{757},
\bfpage{L17}.
\doiurl{https://doi.org/10.1088/2041-8205/757/1/L17}.
\adsurl{2012ApJ...757L..17D}.
\end{barticle}
\endbibitem

\bibitem[\protect\citeauthoryear{{del Toro
  Iniesta}}{2003}]{2003isp..book.....D}
\begin{bbook}
\bauthor{\bsnm{{del Toro Iniesta}}, \binits{J.C.}}:
\byear{2003},
\bbtitle{{Introduction to Spectropolarimetry}},
\bpublisher{Cambridge University Press},
\blocation{Cambridge, UK}.
\adsurl{2003isp..book.....D}.
\end{bbook}
\endbibitem

\bibitem[\protect\citeauthoryear{{del Toro Iniesta} and
  {Collados}}{2000}]{2000ApOpt..39.1637D}
\begin{barticle}
\bauthor{\bsnm{{del Toro Iniesta}}, \binits{J.C.}},
\bauthor{\bsnm{{Collados}}, \binits{M.}}:
\byear{2000},
\batitle{{Optimum Modulation and Demodulation Matrices for Solar Polarimetry}}.
\bjtitle{\applopt}
\bvolume{39},
\bfpage{1637}.
\doiurl{https://doi.org/10.1364/AO.39.001637}.
\adsurl{2000ApOpt..39.1637D}.
\end{barticle}
\endbibitem

\bibitem[\protect\citeauthoryear{{Demidov} and
  {Balthasar}}{2012}]{2012SoPh..276...43D}
\begin{barticle}
\bauthor{\bsnm{{Demidov}}, \binits{M.L.}},
\bauthor{\bsnm{{Balthasar}}, \binits{H.}}:
\byear{2012},
\batitle{{On Multi-Line Spectro-Polarimetric Diagnostics of the Quiet Sun's
  Magnetic Fields. Statistics, Inversion Results and Effects on the SOHO/MDI
  Magnetogram Calibration}}.
\bjtitle{\solphys}
\bvolume{276},
\bfpage{43}.
\doiurl{https://doi.org/10.1007/s11207-011-9863-x}.
\adsurl{2012SoPh..276...43D}.
\end{barticle}
\endbibitem

\bibitem[\protect\citeauthoryear{Duda and Hart}{1972}]{Duda1972UseOT}
\begin{barticle}
\bauthor{\bsnm{Duda}, \binits{R.O.}},
\bauthor{\bsnm{Hart}, \binits{P.E.}}:
\byear{1972},
\batitle{Use of the Hough transformation to detect lines and curves in
  pictures}.
\bjtitle{Commun. Assoc. Comput. Mach. (ACM)}
\bvolume{15},
\bfpage{11}.
\end{barticle}
\endbibitem

\bibitem[\protect\citeauthoryear{{Elmore} et~al.}{1992}]{1992SPIE.1746...22E}
\begin{bchapter}
\bauthor{\bsnm{{Elmore}}, \binits{D.F.}},
\bauthor{\bsnm{{Lites}}, \binits{B.W.}},
\bauthor{\bsnm{{Tomczyk}}, \binits{S.}},
\bauthor{\bsnm{{Skumanich}}, \binits{A.P.}},
\bauthor{\bsnm{{Dunn}}, \binits{R.B.}},
\bauthor{\bsnm{{Schuenke}}, \binits{J.A.}},
\bauthor{\bsnm{{Streander}}, \binits{K.V.}},
\bauthor{\bsnm{{Leach}}, \binits{T.W.}},
\bauthor{\bsnm{{Chambellan}}, \binits{C.W.}},
\bauthor{\bsnm{{Hull}}, \binits{H.K.}}:
\byear{1992},
\bctitle{{The Advanced Stokes Polarimeter - A new instrument for solar magnetic
  field research}}.
In: \beditor{\bsnm{{Goldstein}}, \binits{D.H.}},
\beditor{\bsnm{{Chipman}}, \binits{R.A.}} (eds.)
\bbtitle{Polarization Analysis and Measurement},
\bsertitle{\procspie}
\bseriesno{CS-1746},
\bfpage{22}.
\doiurl{https://doi.org/10.1117/12.138795}.
\adsurl{1992SPIE.1746...22E}.
\end{bchapter}
\endbibitem

\bibitem[\protect\citeauthoryear{{Ferayorni}
  et~al.}{2014}]{2014SPIE.9152E..0ZF}
\begin{bchapter}
\bauthor{\bsnm{{Ferayorni}}, \binits{A.}},
\bauthor{\bsnm{{Beard}}, \binits{A.}},
\bauthor{\bsnm{{Berst}}, \binits{C.}},
\bauthor{\bsnm{{Goodrich}}, \binits{B.}}:
\byear{2014},
\bctitle{{DKIST controls model for synchronization of instrument cameras,
  polarization modulators, and mechanisms}}.
In: \beditor{\bsnm{{Chiozzi}}, \binits{G.}},
\beditor{\bsnm{{Radziwill}}, \binits{N.M.}} (eds.)
\bbtitle{Software and Cyberinfrastructure for Astronomy III},
\bsertitle{\procspie}
\bseriesno{CS-9152},
\bfpage{91520Z}.
\doiurl{https://doi.org/10.1117/12.2057143}.
\adsurl{2014SPIE.9152E..0ZF}.
\end{bchapter}
\endbibitem

\bibitem[\protect\citeauthoryear{{Greisen} and
  {Calabretta}}{2002}]{2002A&A...395.1061G}
\begin{barticle}
\bauthor{\bsnm{{Greisen}}, \binits{E.W.}},
\bauthor{\bsnm{{Calabretta}}, \binits{M.R.}}:
\byear{2002},
\batitle{{Representations of world coordinates in FITS}}.
\bjtitle{\aap}
\bvolume{395},
\bfpage{1061}.
\doiurl{https://doi.org/10.1051/0004-6361:20021326}.
\adsurl{2002A&A...395.1061G}.
\end{barticle}
\endbibitem

\bibitem[\protect\citeauthoryear{{Greisen} et~al.}{2006}]{2006A&A...446..747G}
\begin{barticle}
\bauthor{\bsnm{{Greisen}}, \binits{E.W.}},
\bauthor{\bsnm{{Calabretta}}, \binits{M.R.}},
\bauthor{\bsnm{{Valdes}}, \binits{F.G.}},
\bauthor{\bsnm{{Allen}}, \binits{S.L.}}:
\byear{2006},
\batitle{{Representations of spectral coordinates in FITS}}.
\bjtitle{\aap}
\bvolume{446},
\bfpage{747}.
\doiurl{https://doi.org/10.1051/0004-6361:20053818}.
\adsurl{2006A&A...446..747G}.
\end{barticle}
\endbibitem

\bibitem[\protect\citeauthoryear{{Hanle}}{1924}]{1924ZPhy...30...93H}
\begin{barticle}
\bauthor{\bsnm{{Hanle}}, \binits{W.}}:
\byear{1924},
\batitle{{{\"U}ber magnetische Beeinflussung der Polarisation der
  Resonanzfluoreszenz}}.
\bjtitle{Zeit. Phys.}
\bvolume{30},
\bfpage{93}.
\doiurl{https://doi.org/10.1007/BF01331827}.
\adsurl{1924ZPhy...30...93H}.
\end{barticle}
\endbibitem

\bibitem[\protect\citeauthoryear{{Harrington}}{2021}]{HarringtonGratingPol}
\begin{botherref}
\oauthor{\bsnm{{Harrington}}, \binits{D.M.}}:
2021,
\textit{private communication}.
\end{botherref}
\endbibitem

\bibitem[\protect\citeauthoryear{{Harrington}
  et~al.}{2020}]{2020JATIS...6c8001H}
\begin{barticle}
\bauthor{\bsnm{{Harrington}}, \binits{D.M.}},
\bauthor{\bsnm{{Jaeggli}}, \binits{S.A.}},
\bauthor{\bsnm{{Schad}}, \binits{T.A.}},
\bauthor{\bsnm{{White}}, \binits{A.J.}},
\bauthor{\bsnm{{Sueoka}}, \binits{S.R.}}:
\byear{2020},
\batitle{{Polarization modeling and predictions for Daniel K. Inouye Solar
  Telescope, part 6: fringe mitigation with polycarbonate modulators and
  optical contact calibration retarders}}.
\bjtitle{J. Astron. Tel. Instrum. Syst.}
\bvolume{6},
\bfpage{038001}.
\doiurl{https://doi.org/10.1117/1.JATIS.6.3.038001}.
\adsurl{2020JATIS...6c8001H}.
\end{barticle}
\endbibitem

\bibitem[\protect\citeauthoryear{{Harrington}
  et~al.}{2021}]{2021JATIS...7d8005H}
\begin{barticle}
\bauthor{\bsnm{{Harrington}}, \binits{D.M.}},
\bauthor{\bsnm{{W{\"o}ger}}, \binits{F.}},
\bauthor{\bsnm{{White}}, \binits{A.J.}},
\bauthor{\bsnm{{Sueoka}}, \binits{S.R.}}:
\byear{2021},
\batitle{{Polarization modeling and predictions for DKIST, part 9: flux
  distribution with FIDO}}.
\bjtitle{Journal of Astronomical Telescopes, Instruments, and Systems}
\bvolume{7},
\bfpage{048005}.
\doiurl{https://doi.org/10.1117/1.JATIS.7.4.048005}.
\adsurl{2021JATIS...7d8005H}.
\end{barticle}
\endbibitem

\bibitem[\protect\citeauthoryear{{Hubbard}, {Goodrich}, and
  {Wampler}}{2010}]{2010SPIE.7740E..2RH}
\begin{bchapter}
\bauthor{\bsnm{{Hubbard}}, \binits{J.}},
\bauthor{\bsnm{{Goodrich}}, \binits{B.}},
\bauthor{\bsnm{{Wampler}}, \binits{S.}}:
\byear{2010},
\bctitle{{The ATST base: command-action-response in action}}.
In: \beditor{\bsnm{{Radziwill}}, \binits{N.M.}},
\beditor{\bsnm{{Bridger}}, \binits{A.}} (eds.)
\bbtitle{Software and Cyberinfrastructure for Astronomy},
\bsertitle{\procspie}
\bseriesno{CS-7740},
\bfpage{77402R}.
\doiurl{https://doi.org/10.1117/12.857599}.
\adsurl{2010SPIE.7740E..2RH}.
\end{bchapter}
\endbibitem

\bibitem[\protect\citeauthoryear{{Hunter}}{2007}]{2007CSE.....9...90H}
\begin{barticle}
\bauthor{\bsnm{{Hunter}}, \binits{J.D.}}:
\byear{2007},
\batitle{{Matplotlib: A 2D Graphics Environment}}.
\bjtitle{Comp. Sci. Eng.}
\bvolume{9},
\bfpage{90}.
\doiurl{https://doi.org/10.1109/MCSE.2007.55}.
\adsurl{2007CSE.....9...90H}.
\end{barticle}
\endbibitem

\bibitem[\protect\citeauthoryear{{Iglesias} and
  {Feller}}{2019}]{2019OptEn..58h2417I}
\begin{barticle}
\bauthor{\bsnm{{Iglesias}}, \binits{F.A.}},
\bauthor{\bsnm{{Feller}}, \binits{A.}}:
\byear{2019},
\batitle{{Instrumentation for solar spectropolarimetry: state of the art and
  prospects}}.
\bjtitle{Opt. Eng.}
\bvolume{58},
\bfpage{082417}.
\doiurl{https://doi.org/10.1117/1.OE.58.8.082417}.
\adsurl{2019OptEn..58h2417I}.
\end{barticle}
\endbibitem

\bibitem[\protect\citeauthoryear{{Jaeggli} et~al.}{2010}]{2010MmSAI..81..763J}
\begin{barticle}
\bauthor{\bsnm{{Jaeggli}}, \binits{S.A.}},
\bauthor{\bsnm{{Lin}}, \binits{H.}},
\bauthor{\bsnm{{Mickey}}, \binits{D.L.}},
\bauthor{\bsnm{{Kuhn}}, \binits{J.R.}},
\bauthor{\bsnm{{Hegwer}}, \binits{S.L.}},
\bauthor{\bsnm{{Rimmele}}, \binits{T.R.}},
\bauthor{\bsnm{{Penn}}, \binits{M.J.}}:
\byear{2010},
\batitle{{FIRS: a new instrument for photospheric and chromospheric studies at
  the DST.}}
\bjtitle{\memsai}
\bvolume{81},
\bfpage{763}.
\adsurl{2010MmSAI..81..763J}.
\end{barticle}
\endbibitem

\bibitem[\protect\citeauthoryear{{Johansson} and
  {Goodrich}}{2012}]{2012SPIE.8451E..0JJ}
\begin{bchapter}
\bauthor{\bsnm{{Johansson}}, \binits{E.M.}},
\bauthor{\bsnm{{Goodrich}}, \binits{B.}}:
\byear{2012},
\bctitle{{Simultaneous control of multiple instruments at the Advanced
  Technology Solar Telescope}}.
In: \beditor{\bsnm{{Radziwill}}, \binits{N.M.}},
\beditor{\bsnm{{Chiozzi}}, \binits{G.}} (eds.)
\bbtitle{Software and Cyberinfrastructure for Astronomy II},
\bsertitle{\procspie}
\bseriesno{CS-8451},
\bfpage{84510J}.
\doiurl{https://doi.org/10.1117/12.926441}.
\adsurl{2012SPIE.8451E..0JJ}.
\end{bchapter}
\endbibitem

\bibitem[\protect\citeauthoryear{{Keller} and {Meltzer}}{1966}]{KM66}
\begin{botherref}
\oauthor{\bsnm{{Keller}}, \binits{J.D.}},
\oauthor{\bsnm{{Meltzer}}, \binits{R.J.}}:
1966,
Reflecting diffraction grating for minimizing anomalies.
US Patent 3,237,508.
\end{botherref}
\endbibitem

\bibitem[\protect\citeauthoryear{{Kentischer}
  et~al.}{1998}]{1998A&A...340..569K}
\begin{barticle}
\bauthor{\bsnm{{Kentischer}}, \binits{T.J.}},
\bauthor{\bsnm{{Schmidt}}, \binits{W.}},
\bauthor{\bsnm{{Sigwarth}}, \binits{M.}},
\bauthor{\bsnm{{Uexkuell}}, \binits{M.V.}}:
\byear{1998},
\batitle{{TESOS, a double Fabry-Perot instrument for solar spectroscopy}}.
\bjtitle{\aap}
\bvolume{340},
\bfpage{569}.
\adsurl{1998A&A...340..569K}.
\end{barticle}
\endbibitem

\bibitem[\protect\citeauthoryear{{Kiselman} et~al.}{2011}]{2011A&A...535A..14K}
\begin{barticle}
\bauthor{\bsnm{{Kiselman}}, \binits{D.}},
\bauthor{\bsnm{{Pereira}}, \binits{T.M.D.}},
\bauthor{\bsnm{{Gustafsson}}, \binits{B.}},
\bauthor{\bsnm{{Asplund}}, \binits{M.}},
\bauthor{\bsnm{{Mel{\'e}ndez}}, \binits{J.}},
\bauthor{\bsnm{{Langhans}}, \binits{K.}}:
\byear{2011},
\batitle{{Is the solar spectrum latitude-dependent?. An investigation with
  SST/TRIPPEL}}.
\bjtitle{\aap}
\bvolume{535},
\bfpage{A14}.
\doiurl{https://doi.org/10.1051/0004-6361/201117553}.
\adsurl{2011A&A...535A..14K}.
\end{barticle}
\endbibitem

\bibitem[\protect\citeauthoryear{{Kuckein} et~al.}{2021}]{2021A&A...653A.165K}
\begin{barticle}
\bauthor{\bsnm{{Kuckein}}, \binits{C.}},
\bauthor{\bsnm{{Balthasar}}, \binits{H.}},
\bauthor{\bsnm{{Quintero Noda}}, \binits{C.}},
\bauthor{\bsnm{{Diercke}}, \binits{A.}},
\bauthor{\bsnm{{Trelles Arjona}}, \binits{J.C.}},
\bauthor{\bsnm{{Ruiz Cobo}}, \binits{B.}},
\bauthor{\bsnm{{Felipe}}, \binits{T.}},
\bauthor{\bsnm{{Denker}}, \binits{C.}},
\bauthor{\bsnm{{Verma}}, \binits{M.}},
\bauthor{\bsnm{{Kontogiannis}}, \binits{I.}},
\bauthor{\bsnm{{Sobotka}}, \binits{M.}}:
\byear{2021},
\batitle{{Multiple Stokes I inversions for inferring magnetic fields in the
  spectral range around Cr I 5782 {\r{A}}}}.
\bjtitle{\aap}
\bvolume{653},
\bfpage{A165}.
\doiurl{https://doi.org/10.1051/0004-6361/202140596}.
\adsurl{2021A&A...653A.165K}.
\end{barticle}
\endbibitem

\bibitem[\protect\citeauthoryear{{Kurucz} and
  {Bell}}{1995}]{1995all..book.....K}
\begin{bbook}
\bauthor{\bsnm{{Kurucz}}, \binits{R.L.}},
\bauthor{\bsnm{{Bell}}, \binits{B.}}:
\byear{1995},
\bbtitle{{Atomic line list}}.
\adsurl{1995all..book.....K}.
\end{bbook}
\endbibitem

\bibitem[\protect\citeauthoryear{{Landi Degl'Innocenti} and
  {Landolfi}}{2004}]{2004ASSL..307.....L}
\begin{bbook}
\bauthor{\bsnm{{Landi Degl'Innocenti}}, \binits{E.}},
\bauthor{\bsnm{{Landolfi}}, \binits{M.}}:
\byear{2004},
\bbtitle{{Polarization in Spectral Lines}}
\bseriesno{307}.
\doiurl{https://doi.org/10.1007/978-1-4020-2415-3}.
\adsurl{2004ASSL..307.....L}.
\end{bbook}
\endbibitem

\bibitem[\protect\citeauthoryear{{Leenaarts}}{2020}]{2020LRSP...17....3L}
\begin{barticle}
\bauthor{\bsnm{{Leenaarts}}, \binits{J.}}:
\byear{2020},
\batitle{{Radiation hydrodynamics in simulations of the solar atmosphere}}.
\bjtitle{Liv. Rev. Solar Phys.}
\bvolume{17},
\bfpage{3}.
\doiurl{https://doi.org/10.1007/s41116-020-0024-x}.
\adsurl{2020LRSP...17....3L}.
\end{barticle}
\endbibitem

\bibitem[\protect\citeauthoryear{{Li} et~al.}{1999}]{Li99}
\begin{barticle}
\bauthor{\bsnm{{Li}}, \binits{L.}},
\bauthor{\bsnm{{Chandezon}}, \binits{J.}},
\bauthor{\bsnm{{Granet}}, \binits{G.}},
\bauthor{\bsnm{{Plumey}}, \binits{J.-P.}}:
\byear{1999},
\batitle{{Rigorous and Efficient Grating-Analysis Method Made Easy for Optical
  Engineers}}.
\bjtitle{\applopt}
\bvolume{38},
\bfpage{304}.
\doiurl{https://doi.org/10.1364/AO.38.000304}.
\adsurl{1999ApOpt..38..304L}.
\end{barticle}
\endbibitem

\bibitem[\protect\citeauthoryear{{Lites} et~al.}{2008}]{2008ApJ...672.1237L}
\begin{barticle}
\bauthor{\bsnm{{Lites}}, \binits{B.W.}},
\bauthor{\bsnm{{Kubo}}, \binits{M.}},
\bauthor{\bsnm{{Socas-Navarro}}, \binits{H.}},
\bauthor{\bsnm{{Berger}}, \binits{T.}},
\bauthor{\bsnm{{Frank}}, \binits{Z.}},
\bauthor{\bsnm{{Shine}}, \binits{R.}},
\bauthor{\bsnm{{Tarbell}}, \binits{T.}},
\bauthor{\bsnm{{Title}}, \binits{A.}},
\bauthor{\bsnm{{Ichimoto}}, \binits{K.}},
\bauthor{\bsnm{{Katsukawa}}, \binits{Y.}},
\bauthor{\bsnm{{Tsuneta}}, \binits{S.}},
\bauthor{\bsnm{{Suematsu}}, \binits{Y.}},
\bauthor{\bsnm{{Shimizu}}, \binits{T.}},
\bauthor{\bsnm{{Nagata}}, \binits{S.}}:
\byear{2008},
\batitle{{The Horizontal Magnetic Flux of the Quiet-Sun Internetwork as
  Observed with the Hinode Spectro-Polarimeter}}.
\bjtitle{\apj}
\bvolume{672},
\bfpage{1237}.
\doiurl{https://doi.org/10.1086/522922}.
\adsurl{2008ApJ...672.1237L}.
\end{barticle}
\endbibitem

\bibitem[\protect\citeauthoryear{{Lites} et~al.}{2010}]{2010ApJ...713..450L}
\begin{barticle}
\bauthor{\bsnm{{Lites}}, \binits{B.W.}},
\bauthor{\bsnm{{Casini}}, \binits{R.}},
\bauthor{\bsnm{{Manso Sainz}}, \binits{R.}},
\bauthor{\bsnm{{Jur{\v{c}}{\'a}k}}, \binits{J.}},
\bauthor{\bsnm{{Ichimoto}}, \binits{K.}},
\bauthor{\bsnm{{Ishikawa}}, \binits{R.}},
\bauthor{\bsnm{{Okamoto}}, \binits{T.J.}},
\bauthor{\bsnm{{Tsuneta}}, \binits{S.}},
\bauthor{\bsnm{{Bellot Rubio}}, \binits{L.}}:
\byear{2010},
\batitle{{Scattering Polarization in the Fe I 630 nm Emission Lines at the
  Extreme Limb of the Sun}}.
\bjtitle{\apj}
\bvolume{713},
\bfpage{450}.
\doiurl{https://doi.org/10.1088/0004-637X/713/1/450}.
\adsurl{2010ApJ...713..450L}.
\end{barticle}
\endbibitem

\bibitem[\protect\citeauthoryear{{Lites} et~al.}{2013}]{2013SoPh..283..579L}
\begin{barticle}
\bauthor{\bsnm{{Lites}}, \binits{B.W.}},
\bauthor{\bsnm{{Akin}}, \binits{D.L.}},
\bauthor{\bsnm{{Card}}, \binits{G.}},
\bauthor{\bsnm{{Cruz}}, \binits{T.}},
\bauthor{\bsnm{{Duncan}}, \binits{D.W.}},
\bauthor{\bsnm{{Edwards}}, \binits{C.G.}},
\bauthor{\bsnm{{Elmore}}, \binits{D.F.}},
\bauthor{\bsnm{{Hoffmann}}, \binits{C.}},
\bauthor{\bsnm{{Katsukawa}}, \binits{Y.}},
\bauthor{\bsnm{{Katz}}, \binits{N.}},
\bauthor{\bsnm{{Kubo}}, \binits{M.}},
\bauthor{\bsnm{{Ichimoto}}, \binits{K.}},
\bauthor{\bsnm{{Shimizu}}, \binits{T.}},
\bauthor{\bsnm{{Shine}}, \binits{R.A.}},
\bauthor{\bsnm{{Streander}}, \binits{K.V.}},
\bauthor{\bsnm{{Suematsu}}, \binits{A.}},
\bauthor{\bsnm{{Tarbell}}, \binits{T.D.}},
\bauthor{\bsnm{{Title}}, \binits{A.M.}},
\bauthor{\bsnm{{Tsuneta}}, \binits{S.}}:
\byear{2013},
\batitle{{The Hinode Spectro-Polarimeter}}.
\bjtitle{\solphys}
\bvolume{283},
\bfpage{579}.
\doiurl{https://doi.org/10.1007/s11207-012-0206-3}.
\adsurl{2013SoPh..283..579L}.
\end{barticle}
\endbibitem

\bibitem[\protect\citeauthoryear{Mandel and Wolf}{1995}]{MW95}
\begin{bbook}
\bauthor{\bsnm{Mandel}, \binits{L.}},
\bauthor{\bsnm{Wolf}, \binits{E.}}:
\byear{1995},
\bbtitle{Optical Coherence and Quantum Optics},
\bpublisher{Cambridge University Press},
\blocation{Cambridge, UK}.
\doiurl{https://doi.org/10.1017/CBO9781139644105}.
\end{bbook}
\endbibitem

\bibitem[\protect\citeauthoryear{{Moore}, {Minnaert}, and
  {Houtgast}}{1966}]{1966sst..book.....M}
\begin{bbook}
\bauthor{\bsnm{{Moore}}, \binits{C.E.}},
\bauthor{\bsnm{{Minnaert}}, \binits{M.G.J.}},
\bauthor{\bsnm{{Houtgast}}, \binits{J.}}:
\byear{1966},
\bbtitle{{The solar spectrum 2935 A to 8770 A}}.
\adsurl{1966sst..book.....M}.
\end{bbook}
\endbibitem

\bibitem[\protect\citeauthoryear{{Neckel} and
  {Labs}}{1984}]{1984SoPh...90..205N}
\begin{barticle}
\bauthor{\bsnm{{Neckel}}, \binits{H.}},
\bauthor{\bsnm{{Labs}}, \binits{D.}}:
\byear{1984},
\batitle{{The solar radiation between 3300 and 12500 {\r{A}}}}.
\bjtitle{\solphys}
\bvolume{90},
\bfpage{205}.
\doiurl{https://doi.org/10.1007/BF00173953}.
\adsurl{1984SoPh...90..205N}.
\end{barticle}
\endbibitem

\bibitem[\protect\citeauthoryear{{Perez} and
  {Granger}}{2007}]{2007CSE.....9c..21P}
\begin{barticle}
\bauthor{\bsnm{{Perez}}, \binits{F.}},
\bauthor{\bsnm{{Granger}}, \binits{B.E.}}:
\byear{2007},
\batitle{{IPython: A System for Interactive Scientific Computing}}.
\bjtitle{Comp. Sci. Eng.}
\bvolume{9},
\bfpage{21}.
\doiurl{https://doi.org/10.1109/MCSE.2007.53}.
\adsurl{2007CSE.....9c..21P}.
\end{barticle}
\endbibitem

\bibitem[\protect\citeauthoryear{{Pietarila Graham}, {Danilovic}, and
  {Sch{\"u}ssler}}{2009}]{2009ApJ...693.1728P}
\begin{barticle}
\bauthor{\bsnm{{Pietarila Graham}}, \binits{J.}},
\bauthor{\bsnm{{Danilovic}}, \binits{S.}},
\bauthor{\bsnm{{Sch{\"u}ssler}}, \binits{M.}}:
\byear{2009},
\batitle{{Turbulent Magnetic Fields in the Quiet Sun: Implications of Hinode
  Observations and Small-Scale Dynamo Simulations}}.
\bjtitle{\apj}
\bvolume{693},
\bfpage{1728}.
\doiurl{https://doi.org/10.1088/0004-637X/693/2/1728}.
\adsurl{2009ApJ...693.1728P}.
\end{barticle}
\endbibitem

\bibitem[\protect\citeauthoryear{{Rast} et~al.}{2021}]{2021SoPh..296...70R}
\begin{barticle}
\bauthor{\bsnm{{Rast}}, \binits{M.P.}},
\bauthor{\bsnm{{Bello Gonz{\'a}lez}}, \binits{N.}},
\bauthor{\bsnm{{Bellot Rubio}}, \binits{L.}},
\bauthor{\bsnm{{Cao}}, \binits{W.}},
\bauthor{\bsnm{{Cauzzi}}, \binits{G.}},
\bauthor{\bsnm{{Deluca}}, \binits{E.}},
\bauthor{\bsnm{{de Pontieu}}, \binits{B.}},
\bauthor{\bsnm{{Fletcher}}, \binits{L.}},
\bauthor{\bsnm{{Gibson}}, \binits{S.E.}},
\bauthor{\bsnm{{Judge}}, \binits{P.G.}},
\bauthor{\bsnm{{Katsukawa}}, \binits{Y.}},
\bauthor{\bsnm{{Kazachenko}}, \binits{M.D.}},
\bauthor{\bsnm{{Khomenko}}, \binits{E.}},
\bauthor{\bsnm{{Landi}}, \binits{E.}},
\bauthor{\bsnm{{Mart{\'\i}nez Pillet}}, \binits{V.}},
\bauthor{\bsnm{{Petrie}}, \binits{G.J.D.}},
\bauthor{\bsnm{{Qiu}}, \binits{J.}},
\bauthor{\bsnm{{Rachmeler}}, \binits{L.A.}},
\bauthor{\bsnm{{Rempel}}, \binits{M.}},
\bauthor{\bsnm{{Schmidt}}, \binits{W.}},
\bauthor{\bsnm{{Scullion}}, \binits{E.}},
\bauthor{\bsnm{{Sun}}, \binits{X.}},
\bauthor{\bsnm{{Welsch}}, \binits{B.T.}},
\bauthor{\bsnm{{Andretta}}, \binits{V.}},
\bauthor{\bsnm{{Antolin}}, \binits{P.}},
\bauthor{\bsnm{{Ayres}}, \binits{T.R.}},
\bauthor{\bsnm{{Balasubramaniam}}, \binits{K.S.}},
\bauthor{\bsnm{{Ballai}}, \binits{I.}},
\bauthor{\bsnm{{Berger}}, \binits{T.E.}},
\bauthor{\bsnm{{Bradshaw}}, \binits{S.J.}},
\bauthor{\bsnm{{Campbell}}, \binits{R.J.}},
\bauthor{\bsnm{{Carlsson}}, \binits{M.}},
\bauthor{\bsnm{{Casini}}, \binits{R.}},
\bauthor{\bsnm{{Centeno}}, \binits{R.}},
\bauthor{\bsnm{{Cranmer}}, \binits{S.R.}},
\bauthor{\bsnm{{Criscuoli}}, \binits{S.}},
\bauthor{\bsnm{{Deforest}}, \binits{C.}},
\bauthor{\bsnm{{Deng}}, \binits{Y.}},
\bauthor{\bsnm{{Erd{\'e}lyi}}, \binits{R.}},
\bauthor{\bsnm{{Fedun}}, \binits{V.}},
\bauthor{\bsnm{{Fischer}}, \binits{C.E.}},
\bauthor{\bsnm{{Gonz{\'a}lez Manrique}}, \binits{S.J.}},
\bauthor{\bsnm{{Hahn}}, \binits{M.}},
\bauthor{\bsnm{{Harra}}, \binits{L.}},
\bauthor{\bsnm{{Henriques}}, \binits{V.M.J.}},
\bauthor{\bsnm{{Hurlburt}}, \binits{N.E.}},
\bauthor{\bsnm{{Jaeggli}}, \binits{S.}},
\bauthor{\bsnm{{Jafarzadeh}}, \binits{S.}},
\bauthor{\bsnm{{Jain}}, \binits{R.}},
\bauthor{\bsnm{{Jefferies}}, \binits{S.M.}},
\bauthor{\bsnm{{Keys}}, \binits{P.H.}},
\bauthor{\bsnm{{Kowalski}}, \binits{A.F.}},
\bauthor{\bsnm{{Kuckein}}, \binits{C.}},
\bauthor{\bsnm{{Kuhn}}, \binits{J.R.}},
\bauthor{\bsnm{{Kuridze}}, \binits{D.}},
\bauthor{\bsnm{{Liu}}, \binits{J.}},
\bauthor{\bsnm{{Liu}}, \binits{W.}},
\bauthor{\bsnm{{Longcope}}, \binits{D.}},
\bauthor{\bsnm{{Mathioudakis}}, \binits{M.}},
\bauthor{\bsnm{{McAteer}}, \binits{R.T.J.}},
\bauthor{\bsnm{{McIntosh}}, \binits{S.W.}},
\bauthor{\bsnm{{McKenzie}}, \binits{D.E.}},
\bauthor{\bsnm{{Miralles}}, \binits{M.P.}},
\bauthor{\bsnm{{Morton}}, \binits{R.J.}},
\bauthor{\bsnm{{Muglach}}, \binits{K.}},
\bauthor{\bsnm{{Nelson}}, \binits{C.J.}},
\bauthor{\bsnm{{Panesar}}, \binits{N.K.}},
\bauthor{\bsnm{{Parenti}}, \binits{S.}},
\bauthor{\bsnm{{Parnell}}, \binits{C.E.}},
\bauthor{\bsnm{{Poduval}}, \binits{B.}},
\bauthor{\bsnm{{Reardon}}, \binits{K.P.}},
\bauthor{\bsnm{{Reep}}, \binits{J.W.}},
\bauthor{\bsnm{{Schad}}, \binits{T.A.}},
\bauthor{\bsnm{{Schmit}}, \binits{D.}},
\bauthor{\bsnm{{Sharma}}, \binits{R.}},
\bauthor{\bsnm{{Socas-Navarro}}, \binits{H.}},
\bauthor{\bsnm{{Srivastava}}, \binits{A.K.}},
\bauthor{\bsnm{{Sterling}}, \binits{A.C.}},
\bauthor{\bsnm{{Suematsu}}, \binits{Y.}},
\bauthor{\bsnm{{Tarr}}, \binits{L.A.}},
\bauthor{\bsnm{{Tiwari}}, \binits{S.}},
\bauthor{\bsnm{{Tritschler}}, \binits{A.}},
\bauthor{\bsnm{{Verth}}, \binits{G.}},
\bauthor{\bsnm{{Vourlidas}}, \binits{A.}},
\bauthor{\bsnm{{Wang}}, \binits{H.}},
\bauthor{\bsnm{{Wang}}, \binits{Y.-M.}},
\bauthor{\bsnm{{NSO and DKIST Project}}},
\bauthor{\bsnm{{DKIST Instrument Scientists}}},
\bauthor{\bsnm{{DKIST Science Working Group}}},
\bauthor{\bsnm{{DKIST Critical Science Plan Community}}}:
\byear{2021},
\batitle{{Critical Science Plan for the Daniel K. Inouye Solar Telescope
  (DKIST)}}.
\bjtitle{\solphys}
\bvolume{296},
\bfpage{70}.
\doiurl{https://doi.org/10.1007/s11207-021-01789-2}.
\adsurl{2021SoPh..296...70R}.
\end{barticle}
\endbibitem

\bibitem[\protect\citeauthoryear{{Reardon}, {Uitenbroek}, and
  {Cauzzi}}{2009}]{2009A&A...500.1239R}
\begin{barticle}
\bauthor{\bsnm{{Reardon}}, \binits{K.P.}},
\bauthor{\bsnm{{Uitenbroek}}, \binits{H.}},
\bauthor{\bsnm{{Cauzzi}}, \binits{G.}}:
\byear{2009},
\batitle{{The solar chromosphere at high resolution with IBIS. III. Comparison
  of Ca II K and Ca II 854.2 nm imaging}}.
\bjtitle{\aap}
\bvolume{500},
\bfpage{1239}.
\doiurl{https://doi.org/10.1051/0004-6361/200811223}.
\adsurl{2009A&A...500.1239R}.
\end{barticle}
\endbibitem

\bibitem[\protect\citeauthoryear{{Rimmele} and {the ATST Science Working
  Group}}{2005}]{rimmele2005atst}
\begin{botherref}
\oauthor{\bsnm{{Rimmele}}, \binits{T.}},
\oauthor{\bsnm{{the ATST Science Working Group}}}:
2005,
ATST Science requirements document.
Technical Report SPEC-0001,
NSO/DKIST.
\url{http://nso.edu/wp-content/uploads/2020/10/SPEC-0001_SRD_RevB.pdf}.
\end{botherref}
\endbibitem

\bibitem[\protect\citeauthoryear{{Rimmele} et~al.}{2020}]{2020SoPh..295..172R}
\begin{barticle}
\bauthor{\bsnm{{Rimmele}}, \binits{T.R.}},
\bauthor{\bsnm{{Warner}}, \binits{M.}},
\bauthor{\bsnm{{Keil}}, \binits{S.L.}},
\bauthor{\bsnm{{Goode}}, \binits{P.R.}},
\bauthor{\bsnm{{Kn{\"o}lker}}, \binits{M.}},
\bauthor{\bsnm{{Kuhn}}, \binits{J.R.}},
\bauthor{\bsnm{{Rosner}}, \binits{R.R.}},
\bauthor{\bsnm{{McMullin}}, \binits{J.P.}},
\bauthor{\bsnm{{Casini}}, \binits{R.}},
\bauthor{\bsnm{{Lin}}, \binits{H.}},
\bauthor{\bsnm{{W{\"o}ger}}, \binits{F.}},
\bauthor{\bsnm{{von der L{\"u}he}}, \binits{O.}},
\bauthor{\bsnm{{Tritschler}}, \binits{A.}},
\bauthor{\bsnm{{Davey}}, \binits{A.}},
\bauthor{\bsnm{{de Wijn}}, \binits{A.}},
\bauthor{\bsnm{{Elmore}}, \binits{D.F.}},
\bauthor{\bsnm{{Fehlmann}}, \binits{A.}},
\bauthor{\bsnm{{Harrington}}, \binits{D.M.}},
\bauthor{\bsnm{{Jaeggli}}, \binits{S.A.}},
\bauthor{\bsnm{{Rast}}, \binits{M.P.}},
\bauthor{\bsnm{{Schad}}, \binits{T.A.}},
\bauthor{\bsnm{{Schmidt}}, \binits{W.}},
\bauthor{\bsnm{{Mathioudakis}}, \binits{M.}},
\bauthor{\bsnm{{Mickey}}, \binits{D.L.}},
\bauthor{\bsnm{{Anan}}, \binits{T.}},
\bauthor{\bsnm{{Beck}}, \binits{C.}},
\bauthor{\bsnm{{Marshall}}, \binits{H.K.}},
\bauthor{\bsnm{{Jeffers}}, \binits{P.F.}},
\bauthor{\bsnm{{Oschmann}}, \binits{J.M.}},
\bauthor{\bsnm{{Beard}}, \binits{A.}},
\bauthor{\bsnm{{Berst}}, \binits{D.C.}},
\bauthor{\bsnm{{Cowan}}, \binits{B.A.}},
\bauthor{\bsnm{{Craig}}, \binits{S.C.}},
\bauthor{\bsnm{{Cross}}, \binits{E.}},
\bauthor{\bsnm{{Cummings}}, \binits{B.K.}},
\bauthor{\bsnm{{Donnelly}}, \binits{C.}},
\bauthor{\bsnm{{de Vanssay}}, \binits{J.-B.}},
\bauthor{\bsnm{{Eigenbrot}}, \binits{A.D.}},
\bauthor{\bsnm{{Ferayorni}}, \binits{A.}},
\bauthor{\bsnm{{Foster}}, \binits{C.}},
\bauthor{\bsnm{{Galapon}}, \binits{C.A.}},
\bauthor{\bsnm{{Gedrites}}, \binits{C.}},
\bauthor{\bsnm{{Gonzales}}, \binits{K.}},
\bauthor{\bsnm{{Goodrich}}, \binits{B.D.}},
\bauthor{\bsnm{{Gregory}}, \binits{B.S.}},
\bauthor{\bsnm{{Guzman}}, \binits{S.S.}},
\bauthor{\bsnm{{Guzzo}}, \binits{S.}},
\bauthor{\bsnm{{Hegwer}}, \binits{S.}},
\bauthor{\bsnm{{Hubbard}}, \binits{R.P.}},
\bauthor{\bsnm{{Hubbard}}, \binits{J.R.}},
\bauthor{\bsnm{{Johansson}}, \binits{E.M.}},
\bauthor{\bsnm{{Johnson}}, \binits{L.C.}},
\bauthor{\bsnm{{Liang}}, \binits{C.}},
\bauthor{\bsnm{{Liang}}, \binits{M.}},
\bauthor{\bsnm{{McQuillen}}, \binits{I.}},
\bauthor{\bsnm{{Mayer}}, \binits{C.}},
\bauthor{\bsnm{{Newman}}, \binits{K.}},
\bauthor{\bsnm{{Onodera}}, \binits{B.}},
\bauthor{\bsnm{{Phelps}}, \binits{L.}},
\bauthor{\bsnm{{Puentes}}, \binits{M.M.}},
\bauthor{\bsnm{{Richards}}, \binits{C.}},
\bauthor{\bsnm{{Rimmele}}, \binits{L.M.}},
\bauthor{\bsnm{{Sekulic}}, \binits{P.}},
\bauthor{\bsnm{{Shimko}}, \binits{S.R.}},
\bauthor{\bsnm{{Simison}}, \binits{B.E.}},
\bauthor{\bsnm{{Smith}}, \binits{B.}},
\bauthor{\bsnm{{Starman}}, \binits{E.}},
\bauthor{\bsnm{{Sueoka}}, \binits{S.R.}},
\bauthor{\bsnm{{Summers}}, \binits{R.T.}},
\bauthor{\bsnm{{Szabo}}, \binits{A.}},
\bauthor{\bsnm{{Szabo}}, \binits{L.}},
\bauthor{\bsnm{{Wampler}}, \binits{S.B.}},
\bauthor{\bsnm{{Williams}}, \binits{T.R.}},
\bauthor{\bsnm{{White}}, \binits{C.}}:
\byear{2020},
\batitle{{The Daniel K. Inouye Solar Telescope - Observatory Overview}}.
\bjtitle{\solphys}
\bvolume{295},
\bfpage{172}.
\doiurl{https://doi.org/10.1007/s11207-020-01736-7}.
\adsurl{2020SoPh..295..172R}.
\end{barticle}
\endbibitem

\bibitem[\protect\citeauthoryear{{Scharmer} et~al.}{2003}]{2003SPIE.4853..341S}
\begin{bchapter}
\bauthor{\bsnm{{Scharmer}}, \binits{G.B.}},
\bauthor{\bsnm{{Bjelksjo}}, \binits{K.}},
\bauthor{\bsnm{{Korhonen}}, \binits{T.K.}},
\bauthor{\bsnm{{Lindberg}}, \binits{B.}},
\bauthor{\bsnm{{Petterson}}, \binits{B.}}:
\byear{2003},
\bctitle{{The 1-meter Swedish solar telescope}}.
In: \beditor{\bsnm{{Keil}}, \binits{S.L.}},
\beditor{\bsnm{{Avakyan}}, \binits{S.V.}} (eds.)
\bbtitle{Innovative Telescopes and Instrumentation for Solar Astrophysics},
\bsertitle{\procspie}
\bseriesno{CS-4853},
\bfpage{341}.
\doiurl{https://doi.org/10.1117/12.460377}.
\adsurl{2003SPIE.4853..341S}.
\end{bchapter}
\endbibitem

\bibitem[\protect\citeauthoryear{{Scharmer} et~al.}{2008}]{2008ApJ...689L..69S}
\begin{barticle}
\bauthor{\bsnm{{Scharmer}}, \binits{G.B.}},
\bauthor{\bsnm{{Narayan}}, \binits{G.}},
\bauthor{\bsnm{{Hillberg}}, \binits{T.}},
\bauthor{\bsnm{{de la Cruz Rodriguez}}, \binits{J.}},
\bauthor{\bsnm{{L{\"o}fdahl}}, \binits{M.G.}},
\bauthor{\bsnm{{Kiselman}}, \binits{D.}},
\bauthor{\bsnm{{S{\"u}tterlin}}, \binits{P.}},
\bauthor{\bsnm{{van Noort}}, \binits{M.}},
\bauthor{\bsnm{{Lagg}}, \binits{A.}}:
\byear{2008},
\batitle{{CRISP Spectropolarimetric Imaging of Penumbral Fine Structure}}.
\bjtitle{\apjl}
\bvolume{689},
\bfpage{L69}.
\doiurl{https://doi.org/10.1086/595744}.
\adsurl{2008ApJ...689L..69S}.
\end{barticle}
\endbibitem

\bibitem[\protect\citeauthoryear{{Sheminova}, {Rutten}, and {Rouppe van der
  Voort}}{2005}]{2005A&A...437.1069S}
\begin{barticle}
\bauthor{\bsnm{{Sheminova}}, \binits{V.A.}},
\bauthor{\bsnm{{Rutten}}, \binits{R.J.}},
\bauthor{\bsnm{{Rouppe van der Voort}}, \binits{L.H.M.}}:
\byear{2005},
\batitle{{The wings of Ca II H and K as solar fluxtube diagnostics}}.
\bjtitle{\aap}
\bvolume{437},
\bfpage{1069}.
\doiurl{https://doi.org/10.1051/0004-6361:20042593}.
\adsurl{2005A&A...437.1069S}.
\end{barticle}
\endbibitem

\bibitem[\protect\citeauthoryear{{Socas-Navarro}
  et~al.}{2006}]{2006SoPh..235...55S}
\begin{barticle}
\bauthor{\bsnm{{Socas-Navarro}}, \binits{H.}},
\bauthor{\bsnm{{Elmore}}, \binits{D.}},
\bauthor{\bsnm{{Pietarila}}, \binits{A.}},
\bauthor{\bsnm{{Darnell}}, \binits{A.}},
\bauthor{\bsnm{{Lites}}, \binits{B.W.}},
\bauthor{\bsnm{{Tomczyk}}, \binits{S.}},
\bauthor{\bsnm{{Hegwer}}, \binits{S.}}:
\byear{2006},
\batitle{{Spinor: Visible and Infrared Spectro-Polarimetry at the National
  Solar Observatory}}.
\bjtitle{\solphys}
\bvolume{235},
\bfpage{55}.
\doiurl{https://doi.org/10.1007/s11207-006-0020-x}.
\adsurl{2006SoPh..235...55S}.
\end{barticle}
\endbibitem

\bibitem[\protect\citeauthoryear{{Socas-Navarro}
  et~al.}{2008}]{2008ApJ...674..596S}
\begin{barticle}
\bauthor{\bsnm{{Socas-Navarro}}, \binits{H.}},
\bauthor{\bsnm{{Borrero}}, \binits{J.M.}},
\bauthor{\bsnm{{Asensio Ramos}}, \binits{A.}},
\bauthor{\bsnm{{Collados}}, \binits{M.}},
\bauthor{\bsnm{{Dom{\'\i}nguez Cerde{\~n}a}}, \binits{I.}},
\bauthor{\bsnm{{Khomenko}}, \binits{E.V.}},
\bauthor{\bsnm{{Mart{\'\i}nez Gonz{\'a}lez}}, \binits{M.J.}},
\bauthor{\bsnm{{Mart{\'\i}nez Pillet}}, \binits{V.}},
\bauthor{\bsnm{{Ruiz Cobo}}, \binits{B.}},
\bauthor{\bsnm{{S{\'a}nchez Almeida}}, \binits{J.}}:
\byear{2008},
\batitle{{Multiline Spectropolarimetry of the Quiet Sun at 5250 and 6302
  {\r{A}}}}.
\bjtitle{\apj}
\bvolume{674},
\bfpage{596}.
\doiurl{https://doi.org/10.1086/521418}.
\adsurl{2008ApJ...674..596S}.
\end{barticle}
\endbibitem

\bibitem[\protect\citeauthoryear{{Stenflo}}{1973}]{1973SoPh...32...41S}
\begin{barticle}
\bauthor{\bsnm{{Stenflo}}, \binits{J.O.}}:
\byear{1973},
\batitle{{Magnetic-Field Structure of the Photospheric Network}}.
\bjtitle{\solphys}
\bvolume{32},
\bfpage{41}.
\doiurl{https://doi.org/10.1007/BF00152728}.
\adsurl{1973SoPh...32...41S}.
\end{barticle}
\endbibitem

\bibitem[\protect\citeauthoryear{{Stenflo}}{1982}]{1982SoPh...80..209S}
\begin{barticle}
\bauthor{\bsnm{{Stenflo}}, \binits{J.O.}}:
\byear{1982},
\batitle{{The Hanle Effect and the Diagnostics of Turbulent Magnetic Fields in
  the Solar Atmosphere}}.
\bjtitle{\solphys}
\bvolume{80},
\bfpage{209}.
\doiurl{https://doi.org/10.1007/BF00147969}.
\adsurl{1982SoPh...80..209S}.
\end{barticle}
\endbibitem

\bibitem[\protect\citeauthoryear{{Stenflo} and
  {Keller}}{1997}]{1997A&A...321..927S}
\begin{barticle}
\bauthor{\bsnm{{Stenflo}}, \binits{J.O.}},
\bauthor{\bsnm{{Keller}}, \binits{C.U.}}:
\byear{1997},
\batitle{{The second solar spectrum. A new window for diagnostics of the Sun.}}
\bjtitle{\aap}
\bvolume{321},
\bfpage{927}.
\adsurl{1997A&A...321..927S}.
\end{barticle}
\endbibitem

\bibitem[\protect\citeauthoryear{{Stenflo}, {Keller}, and
  {Gandorfer}}{1998}]{1998A&A...329..319S}
\begin{barticle}
\bauthor{\bsnm{{Stenflo}}, \binits{J.O.}},
\bauthor{\bsnm{{Keller}}, \binits{C.U.}},
\bauthor{\bsnm{{Gandorfer}}, \binits{A.}}:
\byear{1998},
\batitle{{Differential Hanle effect and the spatial variation of turbulent
  magnetic fields on the Sun}}.
\bjtitle{\aap}
\bvolume{329},
\bfpage{319}.
\adsurl{1998A&A...329..319S}.
\end{barticle}
\endbibitem

\bibitem[\protect\citeauthoryear{{Tinbergen}}{2005}]{2005aspo.book.....T}
\begin{bbook}
\bauthor{\bsnm{{Tinbergen}}, \binits{J.}}:
\byear{2005},
\bbtitle{{Astronomical Polarimetry}},
\bpublisher{Cambridge Univ. Press},
\blocation{Cambridge, UK}.
\adsurl{2005aspo.book.....T}.
\end{bbook}
\endbibitem

\bibitem[\protect\citeauthoryear{{Tomczyk} et~al.}{2010}]{2010ApOpt..49.3580T}
\begin{barticle}
\bauthor{\bsnm{{Tomczyk}}, \binits{S.}},
\bauthor{\bsnm{{Casini}}, \binits{R.}},
\bauthor{\bsnm{{de Wijn}}, \binits{A.G.}},
\bauthor{\bsnm{{Nelson}}, \binits{P.G.}}:
\byear{2010},
\batitle{{Wavelength-diverse polarization modulators for Stokes polarimetry}}.
\bjtitle{\ao}
\bvolume{49},
\bfpage{3580}.
\doiurl{https://doi.org/10.1364/AO.49.003580}.
\adsurl{2010ApOpt..49.3580T}.
\end{barticle}
\endbibitem

\bibitem[\protect\citeauthoryear{{Tritschler}
  et~al.}{2002}]{2002SoPh..211...17T}
\begin{barticle}
\bauthor{\bsnm{{Tritschler}}, \binits{A.}},
\bauthor{\bsnm{{Schmidt}}, \binits{W.}},
\bauthor{\bsnm{{Langhans}}, \binits{K.}},
\bauthor{\bsnm{{Kentischer}}, \binits{T.}}:
\byear{2002},
\batitle{{High-resolution solar spectroscopy with TESOS - Upgrade from a double
  to a triple system}}.
\bjtitle{\solphys}
\bvolume{211},
\bfpage{17}.
\doiurl{https://doi.org/10.1023/A:1022459132089}.
\adsurl{2002SoPh..211...17T}.
\end{barticle}
\endbibitem

\bibitem[\protect\citeauthoryear{{Tritschler}
  et~al.}{2016}]{2016AN....337.1064T}
\begin{barticle}
\bauthor{\bsnm{{Tritschler}}, \binits{A.}},
\bauthor{\bsnm{{Rimmele}}, \binits{T.R.}},
\bauthor{\bsnm{{Berukoff}}, \binits{S.}},
\bauthor{\bsnm{{Casini}}, \binits{R.}},
\bauthor{\bsnm{{Kuhn}}, \binits{J.R.}},
\bauthor{\bsnm{{Lin}}, \binits{H.}},
\bauthor{\bsnm{{Rast}}, \binits{M.P.}},
\bauthor{\bsnm{{McMullin}}, \binits{J.P.}},
\bauthor{\bsnm{{Schmidt}}, \binits{W.}},
\bauthor{\bsnm{{W{\"o}ger}}, \binits{F.}},
\bauthor{\bsnm{{DKIST Team}}}:
\byear{2016},
\batitle{{Daniel K. Inouye Solar Telescope: High-resolution observing of the
  dynamic Sun}}.
\bjtitle{Astron. Nach.}
\bvolume{337},
\bfpage{1064}.
\doiurl{https://doi.org/10.1002/asna.201612434}.
\adsurl{2016AN....337.1064T}.
\end{barticle}
\endbibitem

\bibitem[\protect\citeauthoryear{{van der Walt}, {Colbert}, and
  {Varoquaux}}{2011}]{2011CSE....13b..22V}
\begin{barticle}
\bauthor{\bsnm{{van der Walt}}, \binits{S.}},
\bauthor{\bsnm{{Colbert}}, \binits{S.C.}},
\bauthor{\bsnm{{Varoquaux}}, \binits{G.}}:
\byear{2011},
\batitle{{The NumPy Array: A Structure for Efficient Numerical Computation}}.
\bjtitle{Comp. Sci. Eng.}
\bvolume{13},
\bfpage{22}.
\doiurl{https://doi.org/10.1109/MCSE.2011.37}.
\adsurl{2011CSE....13b..22V}.
\end{barticle}
\endbibitem

\bibitem[\protect\citeauthoryear{{von der
  L{\"u}he}}{1998}]{1998NewAR..42..493V}
\begin{barticle}
\bauthor{\bsnm{{von der L{\"u}he}}, \binits{O.}}:
\byear{1998},
\batitle{{High-resolution observations with the German Vacuum Tower Telescope
  on Tenerife}}.
\bjtitle{New Astron. Rev.}
\bvolume{42},
\bfpage{493}.
\doiurl{https://doi.org/10.1016/S1387-6473(98)00060-8}.
\adsurl{1998NewAR..42..493V}.
\end{barticle}
\endbibitem

\bibitem[\protect\citeauthoryear{{W{\"o}ger}
  et~al.}{2021}]{2021SoPh..296..145W}
\begin{barticle}
\bauthor{\bsnm{{W{\"o}ger}}, \binits{F.}},
\bauthor{\bsnm{{Rimmele}}, \binits{T.}},
\bauthor{\bsnm{{Ferayorni}}, \binits{A.}},
\bauthor{\bsnm{{Beard}}, \binits{A.}},
\bauthor{\bsnm{{Gregory}}, \binits{B.S.}},
\bauthor{\bsnm{{Sekulic}}, \binits{P.}},
\bauthor{\bsnm{{Hegwer}}, \binits{S.L.}}:
\byear{2021},
\batitle{{The Daniel K. Inouye Solar Telescope (DKIST)/Visible Broadband Imager
  (VBI)}}.
\bjtitle{\solphys}
\bvolume{296},
\bfpage{145}.
\doiurl{https://doi.org/10.1007/s11207-021-01881-7}.
\adsurl{2021SoPh..296..145W}.
\end{barticle}
\endbibitem

\bibitem[\protect\citeauthoryear{{Zeeman}}{1897a}]{1897ApJ.....5..332Z}
\begin{barticle}
\bauthor{\bsnm{{Zeeman}}, \binits{P.}}:
\byear{1897}a,
\batitle{{On the Influence of Magnetism on the Nature of the Light Emitted by a
  Substance.}}
\bjtitle{\apj}
\bvolume{5},
\bfpage{332}.
\doiurl{https://doi.org/10.1086/140355}.
\adsurl{1897ApJ.....5..332Z}.
\end{barticle}
\endbibitem

\bibitem[\protect\citeauthoryear{{Zeeman}}{1897b}]{1897Natur..55..347Z}
\begin{barticle}
\bauthor{\bsnm{{Zeeman}}, \binits{P.}}:
\byear{1897}b,
\batitle{{The Effect of Magnetisation on the Nature of Light Emitted by a
  Substance}}.
\bjtitle{\nat}
\bvolume{55},
\bfpage{347}.
\doiurl{https://doi.org/10.1038/055347a0}.
\adsurl{1897Natur..55..347Z}.
\end{barticle}
\endbibitem

\end{thebibliography}

\end{article}

\end{document}